\documentclass{nature_plusfigure}
\usepackage{nopageno}

\usepackage{graphicx,float}
\usepackage{subcaption}
\usepackage{amsmath,amssymb,xcolor,url}
\usepackage{multirow}
\usepackage{longtable,tabularx,tablefootnote}
\usepackage{upgreek}
\usepackage{url}
\usepackage{hyperref}
\graphicspath{{./}{figures/}}
\renewcommand{\thefootnote}{\textit{\alph{footnote}}}

\usepackage[none]{hyphenat}
\usepackage{multibib}
\newcites{New}{References}

\newcommand{\mn}{{Mon. Not. R. Astron. Soc.}}
\newcommand{\nar}{{New Astron. Reviews}}
\newcommand{\physrep}{{Phys. Reports}}
\newcommand{\mnras}{\mn}
\newcommand{\aj}{{``Astron. J."}}
\newcommand{\apj}{{Astrophys. J.}}
\newcommand{\apjl}{{Astrophys. J. Lett.}}
\newcommand{\apjs}{{Astrophys. J. Supp.}}
\newcommand{\apss}{{Astrophys. and Space Sc.}}

\newcommand{\aap}{{Astron. Astrophys.}}
\newcommand{\nat}{{Nature}}
\newcommand{\pasj}{{PASJ}}

\newcommand{\pasp}{{Pub. Ast. Soc. Pac.}}
\newcommand{\procspie}{Proc. SPIE}
\newcommand{\ssr}{Space Science Reviews}
\newcommand{\araa}{Annual Review of Astronomy and Astrophysics}

\newcommand{\at}{AT2022tsd}

\usepackage[symbol]{footmisc}
\renewcommand{\thefootnote}{\fnsymbol{footnote}}

\usepackage{cleveref}

\usepackage[none]{hyphenat}
\usepackage{multibib}
\newcites{New}{References}
\usepackage{subcaption} % to add labels to panels

\title{Minutes-duration Optical Flares with Supernova Luminosities}

\author{Anna Y. Q. Ho$^{1}$, Daniel A. Perley$^{2}$, Ping Chen$^{3}$, Steve Schulze$^{4}$, Vik Dhillon$^{5,6}$, Harsh Kumar$^{7}$, Aswin Suresh$^{7}$, Vishwajeet Swain$^{7}$, Michael Bremer$^{8}$, Stephen J. Smartt$^{9,10}$, Joseph P. Anderson$^{11,12}$, G. C. Anupama$^{13}$, Supachai Awiphan$^{14}$, Sudhanshu Barway$^{13}$, Eric C. Bellm$^{15}$, Sagi Ben-Ami$^{3}$, Varun Bhalerao$^{7}$, Thomas de Boer$^{16}$, Thomas G. Brink$^{17}$, Rick Burruss$^{18}$, Poonam Chandra$^{19}$, Ting-Wan Chen$^{20,21}$, Wen-Ping Chen$^{22}$, Jeff Cooke$^{23,24,25}$, Michael W. Coughlin$^{26}$, Kaustav K. Das$^{27}$, Andrew J. Drake$^{27}$, Alexei V. Filippenko$^{17}$, James Freeburn$^{23,24}$, Christoffer Fremling$^{18,28}$, Michael D. Fulton$^{10}$, Avishay Gal-Yam$^{3}$, Llu\'is Galbany$^{29,30}$, Hua Gao$^{16}$, Matthew J. Graham$^{28}$, Mariusz Gromadzki$^{31}$, Claudia P. Guti\'errez$^{30,29}$, K-Ryan Hinds$^{2}$, Cosimo Inserra$^{32}$, Nayana A. J.$^{13}$, Viraj Karambelkar$^{27}$, Mansi M. Kasliwal$^{27}$, Shri Kulkarni$^{27}$, Tom\'as E. M\"uller-Bravo$^{29,30}$, Eugene A. Magnier$^{16}$, Ashish A. Mahabal$^{28,33}$, Thomas Moore$^{10}$, Chow-Choong Ngeow$^{22}$, Matt Nicholl$^{10}$, Eran O. Ofek$^{3}$, Conor M. B. Omand$^{34}$, Francesca Onori$^{35}$, Yen-Chen Pan$^{22}$, Priscila J. Pessi$^{34}$, Glen Petitpas$^{36,37}$, David Polishook$^{38}$, Saran Poshyachinda$^{14}$, Miika Pursiainen$^{39}$, Reed Riddle$^{18}$, Antonio C. Rodriguez$^{27}$, Ben Rusholme$^{40}$, Enrico Segre$^{41}$, Yashvi Sharma$^{27}$, Ken W. Smith$^{10}$, Jesper Sollerman$^{34}$, Shubham Srivastav$^{10}$, Nora Linn Strotjohann$^{3}$, Mark Suhr$^{23,25}$, Dmitry Svinkin$^{42}$, Yanan Wang$^{43,44}$, Philip Wiseman$^{44}$, Avery Wold$^{40}$, Sheng Yang$^{45}$, Yi Yang$^{17}$, Yuhan Yao$^{27}$, David R. Young$^{10}$, WeiKang Zheng$^{17}$
	}

\begin{document}

\maketitle

%\noindent {\bf Affiliations}
\begin{small}
\begin{affiliations}
\label{sec:affiliations}

\item Department of Astronomy, Cornell University, Ithaca, NY 14853, USA
\item Astrophysics Research Institute, Liverpool John Moores University, IC2, Liverpool Science Park, 146 Brownlow Hill, Liverpool L3 5RF, UK
\item Department of Particle Physics and Astrophysics, Weizmann Institute of Science, 234 Herzl St, 76100 Rehovot, Israel
\item The Oskar Klein Centre, Department of Physics, Stockholm University, Albanova University Center, SE 106 91 Stockholm, Sweden
\item Department of Physics and Astronomy, University of Sheffield, Sheffield S3 7RH, UK
\item Instituto de Astrofísica de Canarias, E-38205 La
Laguna, Tenerife, Spain
\item Indian Institute of Technology Bombay, Powai, Mumbai 400076, India
\item Institut de Radioastronomie Millimétrique (IRAM), 300 Rue de la Piscine, F-38406 Saint Martin d’Hères, France
\item Department of Physics, University of Oxford, Denys Wilkinson Building, Keble Road, Oxford OX1 3RH, UK
\item Astrophysics Research Centre, School of Mathematics and Physics, Queen’s University Belfast, Belfast, BT7 1NN, UK
\item European Southern Observatory, Alonso de C\'ordova 3107, Casilla 19, Santiago, Chile
\item Millennium Institute of Astrophysics MAS, Nuncio Monsenor Sotero Sanz 100, Off.104, Providencia, Santiago, Chile
\item Indian Institute of Astrophysics, II Block Koramangala, Bengaluru 560034, India
\item National Astronomical Research Institute of Thailand, 260 Moo 4, Donkaew, Mae Rim, Chiang Mai, 50180, Thailand
\item DIRAC Institute, Department of Astronomy, University of Washington, 3910 15th Avenue NE, Seattle, WA 98195, USA
\item Institute for Astronomy, University of Hawaii, 2680 Woodlawn Drive, Honolulu HI 96822
\item Department of Astronomy, University of California, Berkeley, CA, 94720-3411, USA 
\item Caltech Optical Observatories, California Institute of Technology, Pasadena, CA 91125
\item National Radio Astronomy Observatory, 520 Edgemont Rd, Charlottesville, VA 22903, USA
\item Technische Universit{\"a}t M{\"u}nchen, TUM School of Natural Sciences, Physik-Department, James-Franck-Stra{\ss}e 1, 85748 Garching, Germany
\item Max-Planck-Institut f{\"u}r Astrophysik, Karl-Schwarzschild Stra{\ss}e 1, 85748 Garching, Germany
\item Graduate Institute of Astronomy, National Central University, 300 Jhongda Road, 32001 Jhongli, Taiwan
\item Centre for Astrophysics and Supercomputing, Swinburne University of Technology, Hawthorn, VIC, 3122, Australia
\item Australian Research Council Centre of Excellence for Gravitational Wave Discovery (OzGrav), Australia
\item Australian Research Council Centre of Excellence for All-Sky Astrophysics in 3 Dimensions (ASTRO-3D), Australia
\item School of Physics and Astronomy, University of Minnesota, Minneapolis, Minnesota 55455, USA
\item Cahill Center for Astrophysics, California Institute of Technology, Pasadena, CA 91125, USA
\item Division of Physics, Mathematics and Astronomy, California Institute of Technology, Pasadena, CA 91125, USA
\item Institute of Space Sciences (ICE-CSIC), Campus UAB, Carrer de Can Magrans, s/n, E-08193 Barcelona, Spain
\item Institut d’Estudis Espacials de Catalunya (IEEC), E-08034 Barcelona, Spain.
\item Astronomical Observatory, University of Warsaw, Al. Ujazdowskie 4, 00-478 Warszawa, Poland
\item Cardiff Hub for Astrophysics Research and Technology, School of Physics \& Astronomy, Cardiff University, Queens Buildings, The Parade, Cardiff, CF24 3AA, UK
\item Center for Data Driven Discovery, California Institute of Technology, Pasadena, CA 91125, USA
\item The Oskar Klein Centre, Department of Astronomy, Stockholm University, Albanova University Center, SE 106 91 Stockholm, Sweden
\item INAF-Osservatorio Astronomico d’Abruzzo, via M. Maggini snc, I-64100 Teramo, Italy
\item Kavli Institute for Astrophysics and Space Research, Massachusetts Institute of Technology, Cambridge, MA 02139
\item Harvard-Smithsonian Center for Astrophysics, 60 Garden Street, Cambridge, MA 02138, USA
\item Faculty of Physics, Weizmann Institute of Science, 234 Herzl St, 76100 Rehovot, Israel
\item DTU Space, National Space Institute, Technical University of Denmark, Elektrovej 327, 2800 Kgs. Lyngby, Denmark
\item IPAC, California Institute of Technology, 1200 E. California Blvd, Pasadena, CA 91125, USA
\item Physics Core Facilities, Weizmann Institute of Science, 234 Herzl St, 76100 Rehovot, Israel
\item Ioffe Institute, 26 Politekhnicheskaya, St. Petersburg, 194021, Russia
\item Key Laboratory of Optical Astronomy, National Astronomical Observatories, Chinese Academy of Sciences, Beĳing 100101, China; 
\item School of Physics and Astronomy, University of Southampton, Southampton, SO17 1BJ, UK
\item Henan Academy of Sciences, Zhengzhou 450046, Henan, China

\end{affiliations}
\end{small}

\renewcommand{\thefootnote}{\textit{\alph{footnote}}}
\newcommand{\arcdeg}{\mbox{$^\circ$}}
\newcommand{\arcsec}{$^{\prime\prime}$}
\newcommand{\arcmin}{$^{\prime}$}
\newcommand{\Msol}{\mbox{$M\raisebox{-.6ex}{\odot}$}}
\newcommand\brobor{\smash[b]{\raisebox{0.6\height}{\scalebox{0.5}{\tiny(}}{\mkern-1.5mu\scriptstyle-\mkern-1.5mu}\raisebox{0.6\height}{\scalebox{0.5}{\tiny)}}}}
\captionsetup[table]{name=Table}

\begin{abstract}

In recent years, certain luminous extragalactic optical transients have been observed to last only a few days\cite{Drout2014}. Their short observed duration implies a different powering mechanism from the most common luminous extragalactic transients (supernovae) whose timescale is weeks\cite{Kasen2017}. Some short-duration transients, most notably AT2018cow\cite{Prentice2018}, display blue optical colours and bright radio and X-ray emission\cite{Ho2023}. Several AT2018cow-like transients have shown hints of a long-lived embedded energy source\cite{Margutti2019}, such as X-ray variability\cite{RiveraSandoval2018,Yao2022},
prolonged ultraviolet emission\cite{Chen2023b}, a tentative X-ray quasiperiodic oscillation\cite{Pasham2021,Zhang2022}, and large energies coupled to fast (but subrelativistic) radio-emitting ejecta\cite{Ho2020_Koala,Coppejans2020}. Here we report observations of minutes-duration optical flares in the aftermath of an AT2018cow-like transient, AT2022tsd (the ``Tasmanian Devil''). The flares occur over a period of months, are highly energetic, and are likely nonthermal, implying that they arise from a near-relativistic outflow or jet. Our observations confirm that in some AT2018cow-like transients the embedded energy source is a compact object, either a magnetar or an accreting black hole.

\end{abstract}

\begin{figure*}
\centering

 \begin{subfigure}[t]{1.0\textwidth}
 \centering
     \includegraphics[]{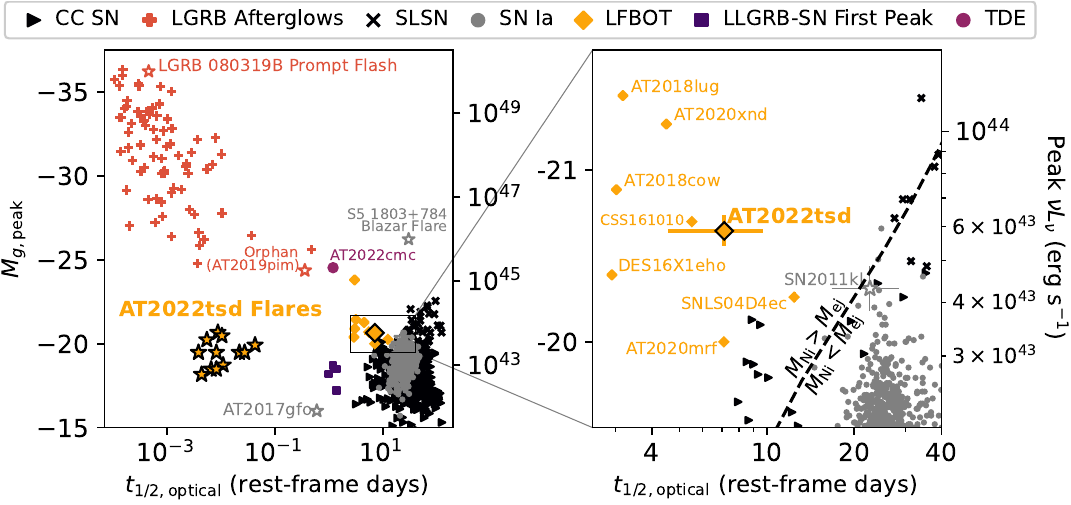}
     \caption*{\textbf{a}}
 \end{subfigure}

 \begin{subfigure}[t]{1.0\textwidth}
 \centering
     \includegraphics[]{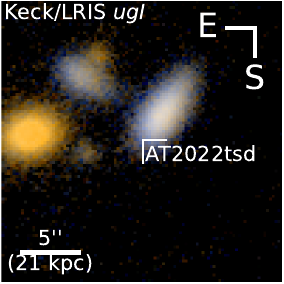}
     \caption*{\textbf{b}}
 \end{subfigure}
\caption{\textbf{\at\ is a fast and luminous extragalactic transient with rapid flares.} \\
\textbf{a} Duration above half-maximum light ($t_{1/2}$) vs. peak absolute magnitude $M$ (or peak luminosity $\nu L_\nu$) of \at, its flares, and other extragalactic optical transients.  
\textbf{b} Keck/LRIS false-colour $u/g/I$ image centred at the position of \at, which is marked.
See Methods section~\ref{sec:data-transient-parameter-space} for additional details and data sources.
}
 \label{fig:transient-parameter-space}
\end{figure*}

In a 30\,s exposure beginning at 11:21:22 on 2022 September 7 (UTC), the Zwicky Transient Facility (ZTF; Methods section~\ref{sec:p48}) detected a new optical transient (internal name ZTF22abftjko) at $r=20.36\pm0.23\,$mag with the position right ascension $\alpha$ = $03^{\rm{h}}20^{\rm{m}}10^{\rm{s}}.873$ and declination $\delta$ = $+08^{\circ} 44' 55''.739$ 
(J2000; uncertainty $0.009''$ from Methods section~\ref{sec:VLA}) as part of its public two-day cadence all-sky survey.
The transient was reported\cite{Munoz-Arancibia2022} to the Transient Name Server by the Automatic Learning for the Rapid Classification of Events (ALeRCE) Alert Broker\cite{Forster2021} and designated AT2022tsd.
Forced photometry on ZTF images (Methods section~\ref{sec:p48}) revealed that the light-curve evolution was faster than that of typical supernovae (Figure~\ref{fig:transient-parameter-space}).
The optical light curve, and the implied high peak luminosity from a nearby (1.4$''$) catalogued galaxy (Methods section~\ref{sec:discovery}, Figure~\ref{fig:transient-parameter-space}), led AT2022tsd to be flagged as a transient of interest as part of ongoing efforts to discover luminous and fast-evolving optical transients (Methods section \ref{sec:discovery}).

\begin{figure*}
 \centering
     \begin{subfigure}[t]{1.0\textwidth}
          \centering        
          \includegraphics[]{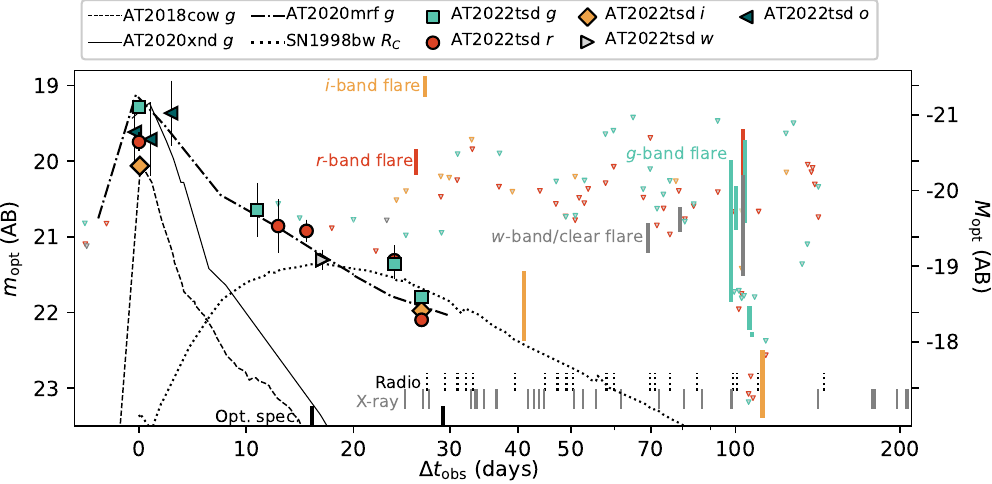}
          \caption*{\textbf{a}}
     \end{subfigure}
     \begin{subfigure}[t]{1.0\textwidth}
          \includegraphics[]{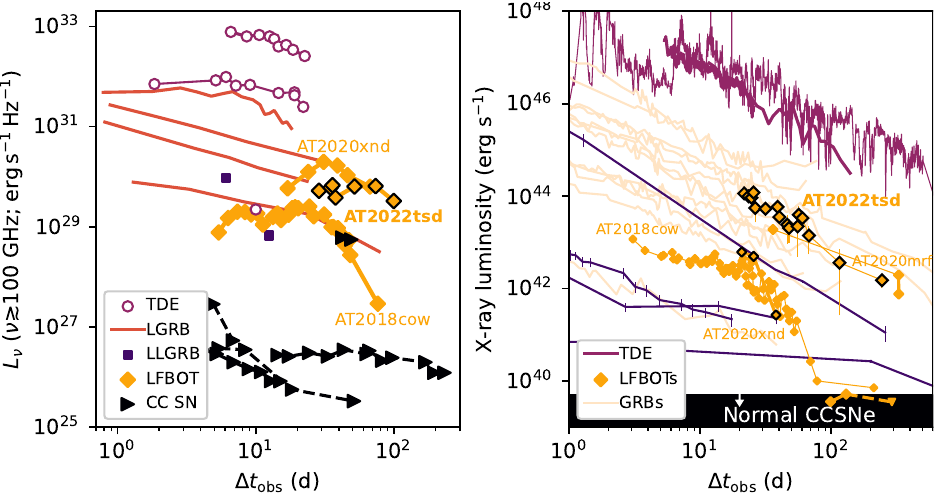}
          \caption*{\textbf{b}}
     \end{subfigure}
\caption{\textbf{\at\ displayed luminous emission across the electromagnetic spectrum.}\\
\textbf{a} Optical light curve of \at\ compared to the luminous fast blue optical transients (LFBOTs) AT2018cow, AT2020xnd, and AT2020mrf, as well as the stripped-envelope SN\,1998bw (associated with GRB\,980425). Vertical bars mark flares, open triangles represent upper limits, and lines along the bottom axis show epochs of radio and X-ray observations as well as optical spectroscopy.\\
\textbf{b} Millimeter-wave and 0.3--10\,keV X-ray light curves of \at\ compared to different classes of extragalactic transients.
Error bars are 1$\sigma$ confidence intervals.
See Methods section~\ref{sec:data-transient-lc} for additional details and data sources. }
 \label{fig:opt-mm-xray-lc}
\end{figure*}

We obtained two spectra of \at\ with the Low Resolution Imaging Spectrometer (LRIS) on the Keck\,I 10-m telescope (\ref{fig:spec}; Methods section \ref{sec:keck}), and measured\cite{Ho2022_Astronote_Keck} a redshift of $z=0.2564\pm0.0003$ (luminosity distance $D_L=1.34\,$Gpc assuming a Planck cosmology\cite{Planck2020}) of the nearby galaxy using prominent narrow host-galaxy emission lines (Methods section \ref{sec:discovery}). 
The optical properties --- the fast light-curve evolution, the implied high peak luminosity ($M_{\mathrm{peak}}=-20.64\pm0.13$ at rest-frame wavelength 5086\AA; Methods section \ref{sec:discovery}), and the lack of prominent spectroscopic features after the transient faded by 2--3 magnitudes --- were unusual for extragalactic transients but similar to AT2018cow, which motivated us to trigger additional multiwavelength observations (Figure~\ref{fig:opt-mm-xray-lc}; Methods section~\ref{sec:multiwavelength-properties}).
We detected luminous radio (decimeter\cite{Ho2022Astronote_radio} to submillimeter) emission that peaked at hundreds of GHz for over a month in the rest frame (Methods section~\ref{sec:VLA}; \ref{fig:radio}),
as well as luminous ($>10^{44}\,$erg\,s$^{-1}$) and steadily fading ($L_X\propto t^{-1.81\pm0.13}$ over nearly 300 days) 0.3--10\,keV X-ray emission\cite{Schulze2022Astronote_xray} well described by a power law with photon index $\Gamma\approx2$ (Methods section~\ref{sec:multiwavelength-properties}, Methods section~\ref{sec:swift}, Figure~\ref{fig:opt-mm-xray-lc}, \ref{fig:xray-lc}).
Although we did not detect clear spectroscopic features from the transient itself, the galaxy alignment is very unlikely to be a coincidence (Methods section \ref{sec:flare-association}), and we conclude that the galaxy is the host of the transient.
The multiwavelength properties of \at\ are most similar to those of AT2018cow-like transients (also referred to as luminous fast blue optical transients or ``LFBOTs''\cite{Metzger2022}), suggesting a common origin (Methods section~\ref{sec:multiwavelength-properties}).

\begin{figure*}
 \begin{subfigure}[t]{\textwidth}
\centering
\includegraphics[width=183mm]{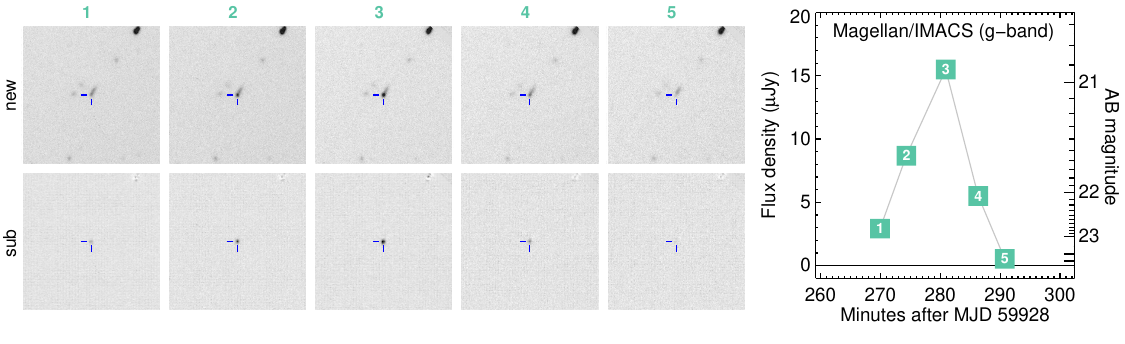}
 \caption*{ \textbf{a}}
\end{subfigure}
\begin{subfigure}[t]{\textwidth}
	\centering
\includegraphics[width=183mm]{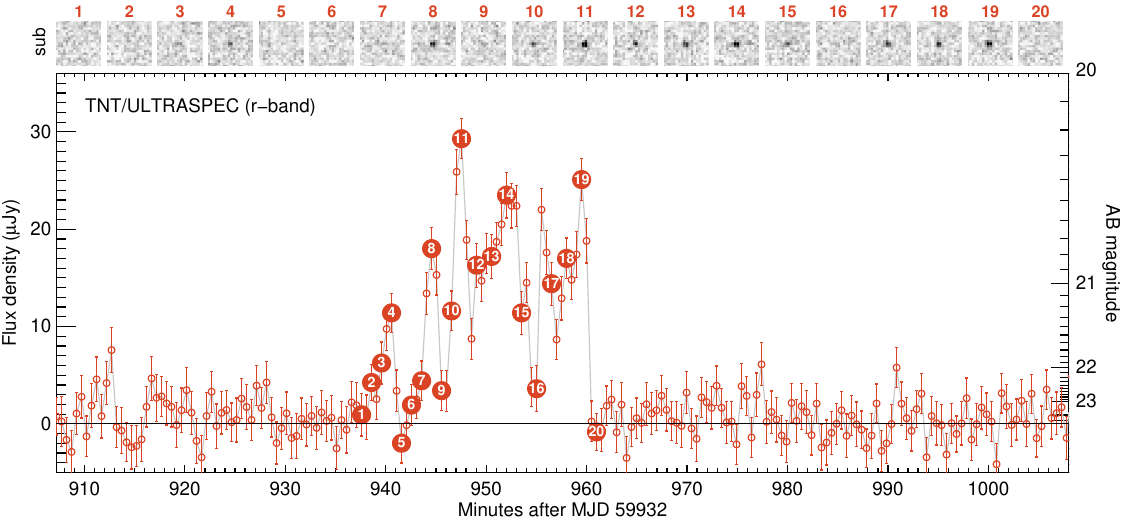}
\caption*{\textbf{b}}
\end{subfigure}
    \caption{\textbf{\at\ exhibited luminous flares lasting tens of minutes.} \textbf{a} Science images (``new''), images with the host galaxy subtracted (``sub''), and the corresponding light curve of a flare detected by Magellan/IMACS at the position of \at\ on 2022 December 15. IMACS observations consisted of five 3\,min-duration exposures. The cutouts show a 45\arcsec\ by 45\arcsec\ region. \textbf{b} Same as \textbf{a} but for a flare detected by ULTRASPEC on 2022 December 19. ULTRASPEC is mounted on the Thai National Telescope, and observations consisted of 30\,s-duration exposures with 15\,msec of dead time between exposures. Error bars are 1$\sigma$ confidence intervals.}
 \label{fig:flare-images-lc}
\end{figure*}

In a photometric optical imaging sequence starting at 04:29:57 on 2022 December 15, 100\,days (observer frame) after the initial transient discovery, we detected\cite{Ho2022_Astronote_Flares} a flare at the position of \at\ across five three-minute Magellan/IMACS $g$-band images (Figure~\ref{fig:flare-images-lc}) that was nearly as bright as the original transient event: $\nu L_\nu \approx 10^{44}\,$erg\,s$^{-1}$ (Figure~\ref{fig:transient-parameter-space}, Figure~\ref{fig:opt-mm-xray-lc}).
Forced photometry on ZTF and Pan-STARRS survey images (Methods section~\ref{sec:panstarrs}) at the position of the transient revealed previous flare detections, as early as 26\,d (observer frame) after the initial transient discovery (Figure~\ref{fig:opt-mm-xray-lc}; \ref{fig:flare-collage}).
Following the IMACS flare detection, we obtained a total of 60\,hr of optical observations of \at\ on 20 different nights, using 13 different telescopes (\ref{tab:flare-searches}). The duration of each sequence ranged from 10\,min to 4.5\,hr. In total we detected at least 14 flares (\ref{fig:flare-collage}). High-cadence ULTRASPEC observations (Methods section~\ref{sec:ultraspec}) revealed flux variations exceeding an order of magnitude on timescales shorter than 20\,s (rest frame; Figure~\ref{fig:flare-images-lc}), and complex temporal profiles that vary between flares (\ref{fig:flare-collage}; Methods section~\ref{sec:flare-characteristics}).
Two different Keck/LRIS observations revealed red flare colours (\ref{fig:flare-collage}; Methods section~\ref{sec:flare-characteristics}):
$u-I=1.41\pm0.31\,$mag, or $\beta=-1.6\pm0.1$ where
$f_\nu \propto \nu^{\beta}$ (corrected for Milky Way extinction but not corrected for host attenuation).

\emph{Chandra} X-ray observations\cite{Matthews2022} (Methods section \ref{sec:chandra}) revealed X-ray variability on timescales of tens of minutes, but no clear high-amplitude flares. 
We detected one definitive optical flare during X-ray monitoring, but no X-ray flare counterpart was detected (\ref{fig:xray-lc}).
In addition, we find no clear periodicity between or within flares in either the optical or X-ray emission (Methods section \ref{sec:flare-characteristics}, \ref{fig:ultraspec-periodogram}, \ref{fig:xray-periodogram}).
We did not identify any high-energy (gamma-ray burst; GRB) counterpart to either the initial LFBOT or the flares (Methods section~\ref{sec:grbsearch}), nor did we identify any similar optical flares in the aftermath of other LFBOTs (Methods section~\ref{sec:flare-search}).
In addition, optical observations of \at\ prior to the first clear flare detection show no significant variability on timescales of minutes (Methods section~\ref{sec:multiwavelength-properties}), implying that there was a longer-duration transient underlying the flares, with a fade rate very similar to that of the LFBOT AT2020mrf\cite{Yao2022} (Figure~\ref{fig:opt-mm-xray-lc}).

To our knowledge, this phenomenon --- minute-timescale optical flares at supernova-like luminosities, with order-of-magnitude amplitude variations, persisting for 100 days --- has no precedent in the literature.
\ref{tab:flaring-classes} lists known classes of objects that exhibit large-amplitude (factor of $\gtrsim10$ times the baseline flux level) flares. Previously observed flaring behaviour was either orders of magnitude less luminous, persisted for only a few minutes, had much longer durations, or was at much higher photon energies.
The fact that these optical flares were observed in the aftermath of an extragalactic transient is even more unusual.

The fast variability timescale of the flares implies an emitting-region radius of $<(9\times10^{11}\,\mathrm{cm}) \Gamma^2$, where $\Gamma$ is the Lorentz factor of the flare-emitting outflow,
and a brightness temperature of $T_B>(2\times10^{10}\,\mathrm{K})\Gamma^{-4}$.
The radius is similar to that inferred from late-time ($\Delta t\approx10^3\,$d) UV observations of AT2018cow\cite{Chen2023b}, and (as in that case) is much smaller than the blackbody radius of the initial LFBOT (Methods section~\ref{sec:lfbot-lc-analysis}).
The high brightness temperature, combined with the red flare colour, implies a nonthermal emission mechanism such as optically thin synchrotron radiation (Methods section \ref{sec:flare-origin}).
The flares are extremely energetic, with $10^{46}$--$10^{47}$\,erg in radiated energy alone per detected flare (not corrected for beaming; \ref{tab:flare-properties}).
In addition, the radiated energy in X-rays during the flaring period exceeds $10^{50}$\,erg.
The timescales, the enormous energetics, the high brightness temperature, and the requirement of optically thin emission for the flares strongly implies that the flare-emitting outflow has at least near-relativistic ($v/c\gtrsim0.6$) velocities (Methods section \ref{sec:flare-origin}),
which reduces the energetics requirements owing to beaming.
However, we have no direct evidence for ultrarelativistic speeds, including a lack of associated detected prompt high-energy emission,
a lack of detected variability at radio wavelengths (Methods section~\ref{sec:ALMA}), and sub-relativistic speeds inferred from a basic equipartition analysis of the radio data (Methods section~\ref{sec:radio-analysis}; Table~\ref{tab:summary}).

We conclude that the flares in \at\ arose from a near-relativistic outflow that was powered by a compact object over a period of 100 days. For the compact object, a supermassive black hole is highly unlikely given the location of \at\ 6\,kpc from the nucleus of a star-forming galaxy (Figure~\ref{fig:transient-parameter-space}, Methods section~\ref{sec:hostgalaxy}) and the rapid timescale of the initial LFBOT. The possible power sources for the outflow are therefore the rotational spindown of a newborn neutron star, or accretion onto a stellar- or intermediate-mass compact object. In the latter case, the compact object could be a newly formed stellar-mass black hole, or, if the process was tidal disruption followed by the formation of an accretion disk, a neutron star, stellar-mass black hole, or intermediate-mass black hole. 

Several models have been proposed to explain LFBOTs\cite{Metzger2022}, and we consider three most likely in light of the newly discovered flares (Methods section~\ref{sec:at2022tsd-origin}): the collapse of a supergiant star\cite{Perley2019,Margutti2019,Quataert2019}, the merger and tidal disruption of a Wolf-Rayet star by a compact object\cite{Metzger2022}, and the tidal disruption of a white dwarf by an intermediate-mass black hole\cite{Kuin2019,Perley2019}.
Accretion processes and jets from systems involving black holes are well known to produce fast and luminous flares,
and explaining \at\ as an analog of observed flares from supermassive black hole tidal disruption events (TDEs) and blazars might be most natural for an intermediate-mass black hole owing to the flare duration and time between flares (tens of minutes to hours).
If \at\ arose from a stellar-mass black hole, the accretion rate would be highly super-Eddington ($10^{5}\,L_\mathrm{Edd}$ for a 10\,$M_\odot$ black hole without relativistic or geometric beaming). Such a rate could be compatible with a merger and tidal disruption scenario\cite{Metzger2022}, and establishing the existence and prevalence of such binary systems is important for understanding the progenitors of merging gravitational-wave sources.
Alternatively, the high accretion rate could arise from the collapse of a supergiant star\cite{Quataert2019} and subsequent formation of an accretion disk; the identification of these systems is a longstanding goal for understanding the conditions that determine whether a star will explode, as well as the formation properties of black holes.
In either picture, the flares could be analogous to the emission observed in GRBs: the timescales are not consistent with external shocks, but could potentially arise from internal shocks.
The lack of detected flares in other LFBOTs could be due to viewing angle: AT2018cow is thought to have been observed close to the plane of the circumburst ``disk'' rather than face-on\cite{Margutti2019,Chen2023b}, and a more on-axis viewing angle for \at\ could also help explain the significantly more luminous X-ray emission (Figure~\ref{fig:opt-mm-xray-lc}).

\noindent {\bf References}

%\bibliographystyle{naturemag}
%\bibliography{references, gcns}

\begin{table}
\begin{center}
\caption{Summary of basic constraints from different emission components.}
\label{tab:summary}
 \begin{tabular}{lcccccc} 
 \hline\hline
 Component & Property & Constraint \\
 \hline
 Prompt Optical & Photospheric radius & $(6.8\pm3.0)\times10^{14}\,$cm \\
 -- & Effective temperature & $(3.3\pm1.8)\times10^{3}\,$K \\
 Optical Flares & Radiated energy & $10^{46}$--$10^{47}\,$erg \\
 -- & Radius (light-crossing time) & $<(9\times10^{11}\,\mathrm{cm})\Gamma^2$ \\
 -- & Brightness temperature & $>(2\times10^{10}\,\mathrm{K})\Gamma^{-4}$ \\
 -- & Equipartition magnetic field strength & $(10^{4}\,\mathrm{G})\Gamma^{-12/7}$ \\
 -- & Equipartition energy & $(10^{43}\,\mathrm{G})\Gamma^{18/7}$ \\
 -- & Velocity & $\gtrsim0.6c$ \\
 Radio & Shock radius (equipartition) & $\gtrsim6\times10^{15}\,$cm \\
 -- & Shock speed (average) & $\gtrsim0.06c$ \\
 -- & Magnetic field strength & $\lesssim6$\,G \\
 -- & Shock energy & $\lesssim3\times10^{48}\,$erg \\
 -- & Ambient density & $\lesssim6\times10^{5}\,$cm$^{-3}$ \\
 X-rays & Radiated energy & $>10^{50}$\,erg \\
 Host Galaxy & Stellar mass & $\log(M/M_\odot)=9.96^{+0.06}_{-0.09}$ \\
 -- & Star-formation rate & $0.55^{+1.36}_{-0.19}\,M_\odot\,$yr$^{-1}$ \\
 \hline
 \end{tabular}
\end{center}
\end{table}

%%%%%%%%%%%%%
% FIGURES 

\clearpage

\begin{methods}

\section{Identification of \at\ and Redshift Measurement}
\label{sec:discovery}

Following the discovery of AT2018cow, we devised and implemented\cite{Ho2023} a filter to discover additional LFBOTs in the ZTF alert stream. Transients are filtered based on age, light-curve timescale (we require duration above half-maximum light $t_{1/2}\lesssim12\,$d\cite{Drout2014}), and peak absolute magnitude (via the best-available host-galaxy redshift estimate).

\at\ was first detected by ZTF (Methods section~\ref{sec:p48}) on 2022 September 7\footnote{UTC dates are used throughout this paper.} as part of its public survey, which images the visible sky in the $g$ and $r$ bands every two nights. Owing to inclement weather and technical issues, the field was next observed on 2022 September 18; on this date, \at\ was not detected with sufficiently high significance (5$\sigma$) for an alert to be generated. On 2022 September 22 ($\Delta t_\mathrm{obs}\footnote{All epochs in this paper are given with respect to the first ZTF detection of \at, which is also the observed peak of the optical light curve.}=15$\,d), forced photometry at the position of \at\ recovered 3$\sigma$ detections on September 18 and September 20, which revealed that the transient had faded by over a magnitude since discovery.
In addition, \at\ was noted to be 1.4\arcsec\ from a catalogued\cite{Beck2021} galaxy in Pan-STARRS (Methods section~\ref{sec:panstarrs}; Figure~\ref{fig:transient-parameter-space}; PSO J050.0451+08.7492; host-galaxy  $g=21.21\pm0.13$\,mag, $r=20.93\pm0.05$\,mag).
The galaxy's photometric redshift\cite{Beck2021} of $z_\mathrm{ph}=0.44\pm0.12$ implied a high peak luminosity (as described later in this section, the true redshift is $z=0.2564$).
The transient met our criteria for fast evolution ($t_{1/2,\mathrm{rise}}<4$\,d and $t_{1/2,\mathrm{fade}}=5.1\pm0.6$\,d) and possible high peak luminosity, so we pursued follow-up spectroscopy.

On 2022 September 23, we obtained a spectrum of \at\ using Keck/LRIS (\ref{fig:spec}; Methods section~\ref{sec:keck}).
\at\ had $r\approx21.5\pm0.2$\,mag at the time, and the slit contained $\sim20\%$ of the host-galaxy flux.
In a 40~min exposure, we detected a blue continuum and a series of prominent host-galaxy emission lines at a consistent redshift. We fit a Gaussian independently to the following emission lines (wavelength given as rest wavelength in air): H$\alpha$ $\lambda$6562.819, H$\beta$ $\lambda$4861.333,  [O~II] $\lambda\lambda$3726.032, 3728.815,
 [O~III] $\lambda\lambda$4958.911, 5006.843, 
  [N~II] $\lambda\lambda$6548.050, 6583.460, and [S~II] $\lambda\lambda$6716.44, 6730.81. 
We measured the redshift by taking the average redshift from the independent fits. The uncertainty in the redshift is set by the small wavelength offset in the line positions between the two Keck spectra (Methods section~\ref{sec:keck}). The result is $z=0.2564\pm0.0003$.
We did not detect any clear spectroscopic features from the transient itself.
Assuming the transient occurred in the galaxy (and the association is highly likely; Methods section~\ref{sec:flare-association}), the implied peak absolute magnitude was $M_{\mathrm{peak}}=-20.64\pm0.13$ at a rest wavelength of 5086\,\AA, accounting for Milky Way extinction ($E_{B-V}=A_V/R_V=0.27\,\mathrm{mag}$, where $R_V=3.1$)\cite{Finkbeiner1999,Schlegel1998,Schlafly2011}. 
To calculate the absolute magnitude, we used the brightest $r$-band detection $m_\mathrm{peak}$ and the following equation, 

\begin{equation}
M_\mathrm{peak} = m_\mathrm{peak} - 5 \log_{10} \left( \frac{D_L}{10\,\mathrm{pc}} \right) + 2.5\,\log_{10} (1+z) \, ,
\end{equation}

\noindent where $D_L$ is the luminosity distance. The duration, absolute magnitude, and blue colours of \at's optical light curve characterise it as an LFBOT (Figure~\ref{fig:transient-parameter-space}). In addition, the lack of prominent spectral features after the transient had faded by over 2\,mag from peak argued against a traditional supernova origin (Methods section~\ref{sec:multiwavelength-properties}). Therefore, we triggered multiwavelength (X-ray through radio) follow-up observations (Methods section~\ref{sec:multiwavelength-properties}) and searched for associated high-energy emission (Methods section~\ref{sec:grbsearch}). Follow-up observations were coordinated using the SkyPortal\cite{vanderWalt2019,Coughlin2023} platform.

\section{Multiwavelength Properties of \at\ Compared to Other Extragalactic Transients}
\label{sec:multiwavelength-properties}

\at\ is only the third LFBOT (after AT2018cow\cite{Prentice2018,Perley2019} and AT2020xnd\cite{Perley2021}) to receive intensive multiwavelength follow-up observations within the first month post-discovery.
Three other LFBOTs (CSS161010\cite{Coppejans2020}, AT2018lug\cite{Ho2020_Koala}, and AT2020mrf\cite{Yao2022}) received their first radio observations only 100\,d post-discovery. MUSSES2020J\cite{Jiang2022} was discovered at $z=1.063$, so follow-up opportunities were limited.
Additional LFBOTs have been identified in archival searches of optical survey data, too late for follow-up observations, such as DES16X1eho\cite{Pursiainen2018} and SNLS04D4ec\cite{Arcavi2016}. 

The peak luminosity ($M_{g,\mathrm{pk}}=-20.64\pm0.13$\,mag), and blue peak colours ($g-r=-0.47\pm0.16$\,mag) of \at's optical light curve are similar to those of AT2018cow\cite{Prentice2018,Perley2019} and AT2020xnd\cite{Perley2021} (Figure~\ref{fig:opt-mm-xray-lc}).
The rise rate is not well constrained ($t_{1/2,\mathrm{rise}}<4\,$d), but is consistent with what was observed for these two objects.
The fade rate ($t_{1/2,\mathrm{fade}}=5.1\pm0.6$\,d, or $\sim0.1\,$mag\,d$^{-1}$) is very similar to that of AT2020mrf\cite{Yao2022}.

Following the Keck/LRIS spectrum on 2022 September 23 ($\Delta t_\mathrm{rest}=13\,$d after peak; Methods section~\ref{sec:discovery}), we obtained a second 40~min Keck/LRIS spectrum on 2022 October 6 ($\Delta t_\mathrm{rest}=23$\,d after peak), when \at\ had $r=22.73\pm0.09\,$mag (\ref{fig:spec}). 
The two Keck spectra are characterised by a blue continuum down to $\sim3000\,$\AA\ in the rest frame, and we do not identify any clear features from the transient itself.\footnote{Despite the lack of distinct transient features, in Methods section~\ref{sec:flare-association} we show that it is highly likely that the transient occurred in the galaxy and is not a foreground object.} A featureless blue continuum so long after peak light, when the light curve has faded by 2--3\,mag, is unusual for extragalactic transients in general\cite{GalYam2017hsn} but has been seen in other LFBOTs.
For example, AT2018cow\cite{Perley2019} exhibited a featureless continuum at $\Delta t=8$\,d, a weak feature at 4850\,\AA\ from $\Delta t=9$\,d to $\Delta t=14$\,d (attributed to He~I $\lambda$4686), and a variety of other lines appearing at 20--30\,d.

The X-ray luminosity of \at\ during the first observation at $\Delta t=20$\,d was $10^{44}\,$erg\,s$^{-1}$, which is similar to that of AT2020mrf\cite{Yao2022} and long-duration gamma-ray burst (LGRB) afterglows; the luminosity is over an order of magnitude greater than that of AT2018cow\cite{RiveraSandoval2018,Margutti2019,Ho2019} or AT2020xnd\cite{Ho2022_AT2020xnd,Bright2022} (Figure~\ref{fig:opt-mm-xray-lc}).
We fit the {\it Swift}/XRT and {\it Chandra}/ACIS detections of \at\ to a power law using the \texttt{curve\_fit} module in \texttt{scipy}, assuming a $t_0$ equal to the first ZTF detection. The best-fit power-law index (\ref{fig:xray-lc}) is $\alpha=-1.81\pm0.13$, where $L_X \propto t^{\alpha}$.
The X-ray light curve of AT2018cow also exhibited a power-law decline near this value\cite{Margutti2019,Ho2019},
which is close to the $t^{-2}$ power law expected for magnetar spindown or accretion under certain conditions\cite{Metzger2022},
and close to $t^{-5/3}$ power law expected for fallback accretion\cite{Phinney1989}.
Binning the {\it Chandra} observations in time revealed variability at the 3$\sigma$ level, with flux variations of factors of a few on timescales of tens of minutes (\ref{fig:xray-lc}). Prolonged rapid X-ray variability was observed in AT2018cow\cite{RiveraSandoval2018,Margutti2019,Ho2019} and AT2020mrf\cite{Yao2022}, and has also been seen in jetted TDEs\cite{Levan2011,Burrows2011,Cenko2012}. An independent analysis of the X-ray data\cite{Matthews2023} found similar values for the luminosity and the temporal power-law index under the assumption of a single power law. 

Unlike the vast majority of extragalactic transients, the spectral energy distribution (SED) of the radio emission from \at\ peaked at hundreds of GHz for months post-discovery (\ref{fig:radio}). To our knowledge, as shown in Figure~\ref{fig:opt-mm-xray-lc}, the only known extragalactic transients with similar behaviour are the LFBOTs AT2018cow\cite{Ho2019} and AT2020xnd\cite{Ho2022_AT2020xnd,Bright2022}.
In addition, the slope of \at's radio SED is significantly shallower than the $f_\nu \propto \nu^{5/2}$ expected from synchrotron self-absorption\cite{Rybicki1986}; the value is closer to $f_\nu \propto \nu^{1}$. A similarly shallow radio SED was observed in AT2018cow\cite{Nayana2021}, and attributed to inhomogeneities in the emitting region or circumburst medium\cite{Nayana2021}.
The shallow spectrum and the persistent peak in the sub-mm bands are more similar to the emission from X-ray binaries (XRBs\cite{Fender2001,Tetarenko2021,Fender2023}) and low-luminosity active galactic nuclei (AGNs) such as Sagittarius A*\cite{Falcke1998} than from explosive transients such as supernovae\cite{Chevalier1998}. 
In the XRB and AGN contexts, the shallow mm-peaking SED is often interpreted as the superposition of self-absorbed components along a continuously powered relativistic jet\cite{Blandford1979}, which we discuss in more detail in Methods section~\ref{sec:at2022tsd-origin}.

\section{Flare Association and Extragalactic Origin}
\label{sec:flare-association}

A hundred days after the discovery of the initial transient event (hereafter referred to as the LFBOT), as part of routine follow-up observations to track the decay of the optical light curve, we detected\cite{Ho2022_Astronote_Flares} a minute-timescale flare at the position of \at\ across five 3~min Magellan/IMACS $g$-band images (Figure~\ref{fig:flare-images-lc}, \ref{fig:flare-collage}, Methods section~\ref{sec:magellan}). A retrospective search of ZTF, Pan-STARRS, and Keck/LRIS data (Methods section~\ref{sec:keck}) revealed additional flare detections as early as $\Delta t_\mathrm{rest}=21\,$d. We searched for detections prior to the LFBOT using ZTF and Pan-STARRS, as might be expected if the flares arose from a foreground Galactic object. There were 190 images obtained by Pan-STARRS going back 3000\,days prior to the LFBOT, with no significant ($>1.4\sigma$) flux excess\cite{Fulton2022}. There were 647 images obtained by ZTF going back 1600\,days prior to the LFBOT, with one image having a $>3\sigma$ flux excess (3.2$\sigma$). The probability of finding at least one image above 3$\sigma$ in 647 images is 60\% (from binomial statistics), so this is not statistically significant. By contrast, of the 65 ZTF exposures obtained from JD~2,459,856.9 to JD~2,459,969.7 (all after the LFBOT), three showed $>3\sigma$ excesses (7.4$\sigma$, 10.1$\sigma$, and 3.5$\sigma$). The probability of finding at least three images above 3$\sigma$ in 65 images is 0.01\%; the probability of finding at least two images above 5$\sigma$ is $1.7\times10^{-8}$. Therefore, it is highly likely that the LFBOT, the multiwavelength (X-ray and radio) emission, and the flares are all associated. 

Given the lack of clear spectroscopic features from the transient itself (Methods section~\ref{sec:multiwavelength-properties}), we considered whether the LFBOT, the multiwavelength emission, and flares could all arise from a foreground source, i.e., whether the proximity to a $z=0.2564$ galaxy could be a chance alignment. 
We note that the Galactic latitude of \at\ is $39.2^\circ$, that there is no counterpart recorded in SIMBAD within 30\arcsec, and that the closest {\it Gaia} DR3 object is 25\arcsec\ away. From our imaging sequence, we estimate that any foreground counterpart would have to be $g \gtrsim 24\,$mag. We considered two classes of events that can resemble LFBOTs owing to their fast blue optical light curves: classical novae and dwarf novae.

Classical novae can produce fast optical light curves and multiwavelength emission\cite{Chomiuk2021}. However, we find a classical nova unlikely for several reasons. First, the peak absolute magnitude of novae ($-5\,$mag to $-10\,$mag\cite{Chomiuk2021}) implies a distance of 1--10\,Mpc for \at, yet there is no nearby galaxy at this position. Second, novae typically show prominent spectral features of H$\alpha$ and other species after maximum optical light\cite{Chomiuk2021}, but the LRIS spectra of \at\ show no such features at $z\approx0$ (\ref{fig:spec}). In addition, the optical to X-ray luminosity ratio of novae is generally $L_\mathrm{opt}/L_X=10^{5}$--$10^{6}$ (for $>1\,$keV X-rays, which typically become detectable one month post-eruption\cite{Chomiuk2021}), whereas in \at\ we observe $L_\mathrm{opt}/L_X\lesssim1$ (\ref{fig:full-sed}).

Dwarf novae, a subclass of cataclysmic variable (CV) outbursts, can also have fast day-timescale blue optical light curves; the optical light curve of \at\ (while sparsely sampled) is similar to that of classified dwarf novae in ZTF's Bright Transient Survey\cite{FremlingBTS,PerleyBTS}. The absolute magnitudes of dwarf novae in quiescence are in the range  8--14\,mag for systems with outburst amplitudes of $\gtrsim4\,$mag\cite{Szkody2021}, implying a distance to \at\ of 1--20\,kpc. At 0.6\,kpc, the X-ray and 10\,GHz radio luminosities of \at\ would be $7\times10^{30}\,$erg\,s$^{-1}$ and $2\times10^{16}$\,erg\,s$^{-1}$\,Hz$^{-1}$, respectively, which is in the observed range for dwarf novae\cite{Polzin2022,Coppejans2020_CV}. 
However, dwarf novae develop prominent spectroscopic features (particularly Balmer lines, He~I, and He~II) after peak light\cite{Morales-Rueda2002,Han2020}. By contrast, we do not see any features at the expected wavelengths of H$\alpha$ or He~I (\ref{fig:spec}). Searching for He~II $\lambda$4686 is complicated by the redshifted [O~II] line, which has a centroid of 4683.5\,\AA\ in the first Keck spectrum and 4686.7\,\AA\ in the second Keck spectrum. 
As discussed in Methods section~\ref{sec:keck}, the shift between the centroids is present in all features at the same level, so is likely due to different slit positions and orientations. In addition, we confirmed that the line-strength ratios are consistent between the two spectra. So, we conclude that we do not detect any contribution from He~II at $z=0$.
Finally, to our knowledge there is no dwarf nova with X-ray emission that decays as a power law for so long after the optical outburst; outside the outburst itself, the X-ray luminosity is typically constant\cite{Fertig2011}.

Another argument disfavouring a CV origin is that the optical flares we observe are very different from the minute-timescale ``flickering'' observed in CVs: CV flickering has much smaller amplitudes (a fraction of a magnitude\cite{Bruch2021}) and a typical flare has blue colours consistent with a hot ($\sim17,000\,$K) blackbody\cite{Bruch2021}.
As a final check, we searched for minute-timescale variability using ZTF light curves of dwarf novae. 
We employed the ZTF Bright Transient Survey\cite{FremlingBTS} Sample Explorer\cite{PerleyBTS} to identify 182 CVs with peak apparent brightness fainter than 18\,mag and that do not have bright quiescent counterparts. Note that BTS requires transients to have a Galactic latitude of at least $7^{\circ}$.
For each object, we retrieved a forced-photometry light curve from the IPAC service (Methods section~\ref{sec:p48}), from March 2018 (the start of the survey) until the end of 2022. For each CV, we searched each night of observations for pairs of subtractions in the same filter and based on the same reference stack.
To count as a flare, a pair of detections had to have a flux change exceeding a factor of 10, and the flux difference had to be significant ($>3\sigma$).
We identified eight candidate flares from six distinct objects. Visual inspection of the science images and difference images revealed that the brightness variations were due to cosmic rays (two images; ZTF18abyxlas and ZTF20acufmrl), a likely ``ghost'' (an artifact of internal reflection, with significant drift from image to image; three images of ZTF18acbwkqu), and a streak (one image; ZTF19abljehr). An additional image (of ZTF19abylcik) had a data-quality flag (\texttt{infobitssci}) and visual inspection showed a positive residual at the location of a nearby star, in addition to a positive residual at the location of the CV; the flag, together with the by-eye assessment of the subtraction, suggest that this positive residual was also an artifact. The remaining object (ZTF18acxhfkq) had a bright point-like counterpart in PS1, the light curve revealed highly significant negative flux values, and visual inspection of the images showed a low significance for the positive residuals; thus, the variability is not robust.  
Therefore, we find that among dwarf novae there is no precedent for flaring with the timescale and amplitude seen in \at.

We conclude that if \at\ is a foreground source, it would be a highly exotic object, and it would be unlikely for such an unusual stellar system to be aligned with a galaxy (Figure~\ref{fig:transient-parameter-space}) whose redshift implies LFBOT-like optical, X-ray, and radio luminosities.
For a crude estimate of the probability of chance alignment, we used the COSMOS photometric redshift catalogue\cite{Ilbert2008} to estimate the density of galaxies brighter than 22\,mag with $0.1 \leq z \leq 0.3$. We found that the number density is $\sim 1000$\,deg$^{-2}$.
A spatial offset of 6\,kpc corresponds to 3\arcsec\ for $z=0.1$, so for each galaxy a transient would have to be within a 30-square-arcsecond region to be considered aligned. For 1000 galaxies in a square-degree region, that gives a covering fraction of 0.002 in which a transient could be considered aligned with a galaxy at the appropriate redshift.
During the second year of ZTF, 372 CV candidates were discovered\cite{Szkody2021}, most of which were dwarf novae; we estimate a rate of 400 per year in the 15,000\,deg$^{2}$ of the ZTF public survey, or 0.02\,deg$^{-2}$\,yr$^{-1}$. So, in a given year, the chance of detecting an uncatalogued dwarf nova aligned with a $z=0.1$--0.3 galaxy is $\sim 4\times10^{-5}$; over the course of five years in ZTF, we estimate $2\times10^{-4}$. Assuming the flaring in \at\ occurs in 1/100 dwarf novae, we find $2\times10^{-6}$.
So, we conclude that the most likely explanation is that \at\ is extragalactic.

\section{Flare Observational Characteristics}
\label{sec:flare-characteristics}

After the discovery of the Magellan/IMACS flare (Figure~\ref{fig:flare-images-lc}), we searched for additional flares with 13 different instruments (\ref{tab:flare-searches}). Here we summarise the observed properties of the flares we detected, which are also listed in \ref{tab:flare-properties}. For each flare, we measured the time interval in which 90\% of the flux was detected ($T_{90}$). The value of $T_{90}$ ranged from $\sim10$\,min (the LT flare, and the small ULTRASPEC $g$-band flare prior to the large flaring episode; \ref{fig:flare-collage}) to 80\,min (the large ULTRASPEC $g$-band flare; \ref{fig:flare-collage}). 

The observed optical flares (Figure~\ref{fig:opt-mm-xray-lc}, \ref{fig:flare-collage}) exhibit a variety of morphologies. The ULTRASPEC $g$-band flare (\ref{fig:flare-collage}) showed a multihour flaring ``episode'' with two prominent peaks superimposed on an exponential decline, as well as a short precursor flare lasting just a few minutes. The ULTRASPEC $r$-band flare (Figure~\ref{fig:flare-images-lc}) was more erratic, with an abrupt turnoff rather than an exponential decline. A Lomb-Scargle periodogram\cite{Lomb1976,Scargle1982} revealed no significant periodicity in the ULTRASPEC light curves (\ref{fig:ultraspec-periodogram}), nor in the X-ray observations (\ref{fig:xray-periodogram}). 

The ULTRASPEC $r$-band flare shows strong variability (Figure~\ref{fig:flare-images-lc}), with order-of-magnitude changes in flux on timescales much shorter than the overall duration of the outburst. The time to change by order unity, $\delta t$, is limited by the 30\,s cadence of the observations. The ratio of this variability time to the overall duration of the burst is therefore $\delta t/T<2\times10^{-2}$. For the ULTRASPEC $g$-band flare (\ref{fig:flare-collage}), the time to change by a factor of order unity is resolved by the individual observations, and is approximately a few minutes. We find $\delta t/T<4\times10^{-2}$.

From the Keck/LRIS observations (\ref{fig:flare-collage}), we can measure the optical-flare colour. The $g+I$ flare detection on 2022 October 19 gives $f_\nu \propto \nu^{-0.45\pm0.01}$ at the start of the sequence, with a trend toward bluer colours over the next 20\,min. The colour evolution may be due to an increasing contribution from the underlying blue transient, rather than a colour change inherent to the flare mechanism.
The $u+I$ flare detection on 2022 December 29 gives $f_\nu \propto \nu^{-1.6\pm0.1}$. There was only one clear detection in both bands during the $u+I$ sequence, so we cannot draw conclusions about the colour evolution using the $u+I$ observations. 

We have simultaneous X-ray and optical observations during one flare (\ref{fig:xray-lc}). We detected an optical flare with LRIS at 10:10 on 2022-12-19, with significant emission lasting for $\sim 20$\,min. We have no constraint on the start time of the optical flare (the previous optical observation ended three days prior). There is no obvious X-ray excess at the time of observed optical peak. The average X-ray luminosity during this epoch is $10^{43}\,$erg\,s$^{-1}$, while the peak observed optical luminosity is $\sim 10^{42}$\,erg\,s$^{-1}$.
Adopting $10^{17}\,$Hz for the X-ray frequency and $10^{14}\,$Hz for the optical frequency, we rule out an optical to X-ray spectral index shallower than $\beta=-4/3$ where $L_\nu = \nu^{\beta}$. 

We estimated the flare duty cycle for different limiting-magnitude thresholds, assuming a Poisson distribution for the likelihood of detecting a flare in any given time interval.
We performed the calculation using all images in the MJD range 59856.4--59942.4 (from the first to last flare detection) except the PS1 $w$-band images, because the wide filter makes it difficult to convert the measurement to a specific filter.
We converted each detection to its estimated $g$-band value, using the measured colour of the flares.
For each threshold, \ref{tab:flare-stats} gives the total number of exposures above that threshold (the number of exposures in which a flare brighter than the threshold could have been detected), the total exposure time of those exposures, and the fraction of time in which a flare was detected. 

To estimate the uncertainty in the duty cycle, we performed a simulation as follows. We adopted a range of flare durations for each threshold (10--20\,min for 21\,mag, and 1\,min to 3\,hr for 22.5\,mag and 24\,mag), based on what we observed. For each choice of flare duration and average flare frequency, we simulated 1000 sets of flare start times from one day prior to our earliest detected flare to one day after our last detected flare. We calculated what the observed duty cycle would have been, and discarded values of average flare frequency that resulted in $<2.5\%$ of the 1000 trials being above or below our true observed value.  As shown in \ref{tab:flare-stats}, bright ($<21$\,mag) flares have a maximum allowed duty cycle of 10\%. Constraints are weak for fainter ($\gtrsim24$\,mag) flares owing to limited observations.  

Finally, we searched for periodicity in the flare occurrence times. The longest continuously observed interval without a flare detection was 3\,hr (ULTRASPEC $r$-band; \ref{fig:flare-collage}). The shortest continuously observed interval between two flares was also several hours (ULTRACAM and KP84), or possibly half an hour if the two flares observed by ULTRASPEC in $g$ were truly distinct. We folded the optical observations by periods between 3\,hr and 1\,d, in 1\,s steps.
We did not identify any clear period that aligned the flares, particularly taking into account our nondetections.
Several short periods (3.35\,hr, 3.7\,hr) aligned the flares to a 2\,hr window, and slightly longer periods (5.0\,hr, 5.1\,hr) to within a $\sim2.7$\,hr window.

\section{Limit on an Associated GRB}
\label{sec:grbsearch}

We searched for a GRB counterpart in the 3.0\,d between the last ZTF nondetection (4 Sep.; JD 2,459,826.9464) and the first ZTF detection of \at. We did not identify any burst consistent with the time and position of \at\ in the GCN archive or the {\it Fermi} burst catalogue. Konus-Wind was taking data throughout this interval, but detected no events consistent with the \at\ position. 
We adopt a 10\,keV -- 10\,MeV fluence and peak flux threshold of few $\times10^{-7}$\,erg\,cm$^{-2}$ and few $\times10^{-7}$\,erg\,cm$^{-2}$\,s$^{-1}$,
respectively (which correspond to the dimmer end of GRBs detected by Konus-Wind in the waiting mode\cite{Tsvetkova2021}),
giving upper limits of $E_{\gamma,\mathrm{iso}} <$ few $\times 10^{49}$\,erg and $L_{\gamma,\mathrm{iso}} <$ few $\times10^{49}$\,erg\,s$^{-1}$. 
These limits rule out an on-axis classical long-duration GRB, but not an off-axis or low-luminosity GRB\cite{Cano2017}.
In addition, these limits are for sources with typical GRB prompt emission timescales; we cannot rule out an ultra-long duration GRB such as Swift J1644+57.
We also searched for GRBs consistent with the position of \at\ between the first ZTF detection and 2023-04-27, but found no reliably associated bursts.

\section{Search for Flares in Other LFBOTs}
\label{sec:flare-search}

The discovery of flares in the aftermath of \at\ (Methods section~\ref{sec:flare-association}) raises the question of whether there could have been flares associated with other LFBOTs. 
Over the years 2018--2022, six LFBOTs were identified in addition to \at: AT2018cow\cite{Prentice2018}, AT2018lug\cite{Ho2020_Koala}, AT2020xnd\cite{Perley2021}, AT2021ahuo, AT2022abfc\cite{Ho2022_AT2022abfc},  
and AT2020mrf\cite{Yao2022}. 
We performed forced photometry on ZTF images at the position of all six objects, with a start date of JD~2,458,194.5 (17 March 2018) and an end date of JD~2,459,944.5 (31 December 2022), identifying no significant flares.
However, for most objects the nominal ZTF survey data cannot be used to rule out flaring with the duty cycle of \at.
There were two tentative 3$\sigma$ detections in the $r$ band, 60\,d after the discovery of AT2021ahuo. However, with only two detections at low significance, it is difficult to determine if they are true flares.
AT2018cow was observed intensely by a variety of optical telescopes during the 80\,d post-discovery\cite{Perley2019}. At the distance of AT2018cow, the threshold of 24.0\,mag for \at\ corresponds to a threshold of 17.4\,mag for AT2018cow. We consider flares of duration 10\,min and 1\,hr. The 964 photometric points can be binned into 497 blocks of 10\,min each, or 257 blocks of 1\,hr. We rule out flares as bright as 17.4\,mag (corresponding to $M=-16.6\,$mag) for all images. We find an upper limit on the duty cycle of 10\,min and 1\,hr flares to be 0.7\% and 1.4\%, respectively (95\% confidence), lower than the 3\% bound for the equivalent threshold in \at. Therefore, we conclude that AT2018cow did not exhibit flaring behaviour with the same duty cycle as \at.

We also performed forced photometry at the position of the LFBOT CSS161010\cite{Coppejans2020} ($z=0.033$). 
We used the online Asteroid Terrestrial-impact Last Alert System (ATLAS; Methods section~\ref{sec:atlas})
forced-photometry service (Methods section~\ref{sec:atlas}) to identify 480 images within 600\,d after the transient. There is no $\geq5\sigma$ detection after the original transient. At the distance of CSS161010, the threshold for 24.0\,mag for \at\ corresponds to 19.2\,mag. The number of images that are sufficiently sensitive, binned by hour, between 20\,d and 100\,d after the transient, is only 8. Therefore, we cannot exclude flaring with a duty cycle identical to that of \at.
ZTF forced photometry also did not identify any significant flares. A 4$\sigma$ ``detection'' turned out upon visual inspection to arise from an image artifact (streak). 

\section{Physical Origin of \at's Flares}
\label{sec:flare-origin}

In this section, we use the observational characteristics of the \at\ flares (Methods section~\ref{sec:flare-characteristics}) to set constraints on their physical origin.

The lowest frequency with clear detected variability is the optical band, so we use this to estimate the brightness temperature of the flares.
From the ULTRASPEC $r$-band observations, the shortest timescale of variability we resolve is $\delta t_\mathrm{obs} = 30\,\mathrm{s}$, setting a limit on the emission-region radius $R$ of $R<\Gamma^2 c \delta t_\mathrm{obs} \approx (9\times10^{11}\,\mathrm{cm}) \Gamma^2$, where $\Gamma$ is the Lorentz factor of the outflow. 
The source angular radius is therefore $d\theta < 7 \times 10^{-5} \Gamma^2 \,\mu$as. Taking the intensity of the brightest ULTRASPEC flare detection (65\,$\mu$Jy in the rest frame), we find $T_B > I_\nu c^2/(2 k\nu^2) \approx 2\times10^{10} \Gamma^{-4} \,$K.
For reasonable values of the Lorentz factor,
the limiting blackbody temperature would result in very blue optical emission ($f_\nu\propto\nu^2$), yet all of the observed optical-flare colours are significantly redder.
Therefore, we consider the emission more likely to be nonthermal.
In addition, the value of $T_B=2\times10^{10}\,$K is very close to the equipartition brightness temperature limit\cite{Readhead1994} of $10^{11}$\,K, suggesting that the outflow is at least close to relativistic. 

Optically thin synchrotron radiation is a possible candidate for the nonthermal flare emission. The flux density from a population of synchrotron-emitting electrons in a power-law energy distribution $N(E)dE = \kappa E^{-p} dE$, where $N(E)dE$ is the number density of electrons in the energy interval $E$ to $E+dE$ in units of $\mathrm{cm}^{-3}\,\mathrm{erg}^{-1}$, is\cite{Longair2011}

\begin{equation}
\label{eq:synchrotron-flux}
    J(\nu) = 2.344\times10^{-37}\, a(p) (10^4\,B)^{(p+1)/2} \kappa \left( \frac{1.253\times10^{37}}{\nu} \right)^{(p-1)/2}\,\mathrm{erg}\,\mathrm{s}^{-1}\,\mathrm{cm}^{-3}\,\mathrm{Hz}^{-1}\,,
\end{equation}

\noindent where $B$ is the magnetic field strength and $\nu$ is the observed frequency. The optical depth to synchrotron self-absorption at a given frequency is $\tau_\nu = \chi_\nu R$, where $R$ is the line-of-sight path length and the absorption coefficient $\chi_\nu$ is

\begin{equation}
\label{eq:synchrotron-tau}
\chi_\nu = 3.354\times10^{-24}\, \kappa (10^4\,B)^{(p+2)/2} (3.54\times10^{18})^p b(p) \nu^{-(p+4)/2}\,\mathrm{cm}^{-1}\, .
\end{equation}

We assume $p=2.5$, which corresponds to\cite{Longair2011} $a(p)=0.359$ and $b(p)=0.244$. Adopting the observed peak flux density of the Keck/LRIS $u+I$ flare, and the inferred size from the variability timescale $R=(1.8\times10^{12}\,\mathrm{cm})\Gamma^2$, we find that the frequency at which the optical depth is unity (the synchrotron self-absorption frequency $\nu_\mathrm{SSA}$) is

\begin{equation}
\label{eq:ssa}
\nu_\mathrm{SSA} = (2\times10^{14}\,\mathrm{Hz}) \left(\frac{B}{\mathrm{G}} \right)^{0.14} \Gamma^{-1.14}\, .
\end{equation}

\noindent Therefore, given the observed characteristics of the \at\ flares, the inferred synchrotron self-absorption frequency is very close to the optical band, consistent with our observation of optically thin emission.
If the flares are synchrotron emission, we can  
estimate the equipartition energy $U_\mathrm{eq}$ and magnetic field strength $B_\mathrm{eq}$. The latter is\cite{Moffet1975}

\begin{equation}
B_\mathrm{eq} = \left( \frac{8\pi A g(\alpha) L}{V} \right)^{2/7}\, ,
\end{equation}

\noindent where $A=1.586\times10^{12}$ in cgs units, $L$ is the luminosity, $V$ is the volume of the synchrotron-emitting electrons, and $g(\alpha)$ is a function of the spectral index $\alpha$ (defined as $f_\nu\propto \nu^{\alpha}$) and frequency range ($\nu_1$ to $\nu_2$) for the power law:

\begin{equation}
    g(\alpha) = \frac{2\alpha+2}{2\alpha+1} 
    \left[ \frac{\nu_2^{\alpha+1/2}-\nu_1^{\alpha+1/2}}{\nu_2^{\alpha+1}-\nu_1^{\alpha+1}} \right].
\end{equation}

From the Keck/LRIS flares we have $L=10^{43}\,$erg\,s$^{-1}$ and $\alpha=-1.6$.
We assume that the power law extends from 
$10^{13}$\,Hz to $10^{15}$\,Hz.
From the variability timescale, we have a radius of the synchrotron-emitting electron sphere of $(9\times10^{11}\,\mathrm{cm})\Gamma^2$.
Taken together, we find $B_\mathrm{eq}\approx(10^{4}\,\mathrm{G})\Gamma^{-12/7}$, which is relatively insensitive to our choices of $\nu_1$, $\nu_2$, and $\alpha$.

Next, we estimate the equipartition energy,

\begin{equation}
    U_\mathrm{eq} = 2 \frac{VB^2}{8\pi}.
\end{equation}

We find  $U_\mathrm{eq}\approx(10^{43}\,$erg)$\Gamma^{18/7}$. Our estimated value of $U_\mathrm{eq}$ can be reconciled with the radiated flare energy in one of two ways: the flare-emitting outflow could be ultrarelativistic, or the electrons could be fast-cooling. Both scenarios are plausible; the observed spectral index ($f_\nu\propto\nu^{-1.6\pm0.1}$) is relatively steep, and the high $B_\mathrm{eq}$ implies a synchrotron cooling time that is much shorter than the dynamical time of the system. %However, it could also be the case that the outflow is ultrarelativistic.

Given the values above, we can estimate the Lorentz factor of the particles emitting in the optical band, $\gamma_e$. The characteristic frequency of those electrons $\nu_e$ is related to the gyrofrequency $\nu_g=q_e B/(2\pi m_e c)$ as $\nu_e = \gamma_e^2 \nu_g$.
At $10^{15}$\,Hz we find $\gamma_e\approx10^{2}\, \Gamma^{6/7}$.

Finally, we estimate the velocity of the flare-emitting outflow. Assuming that the kinetic energy of the outflow in \at\ is on the order of the observed optical flare luminosity, we have $L_\mathrm{opt} \approx 10^{44}\,$erg\,s$^{-1} \approx \eta \dot{M}v^2$ (for the brightest flares), where $\dot{M}$ and $v$ are the mass-loss rate and velocity of the outflow (respectively), and $\eta$ is the efficiency of converting kinetic energy to radiation. In this case, the observed nonthermal emission must arise from a radius that is larger than the Thomson scattering photosphere. In the observer frame, the optical depth to Thomson scattering is

\begin{equation}
\tau = n_e \sigma_T R\, ,
\end{equation}

\noindent where $\sigma_T$ is the scattering cross-section, $R$ is the depth into the outflow (assumed to be comparable to the radius of the outflow), and

\begin{equation}
    n_e = \frac{\dot{M}}{4\pi m_p R^2 v}\, .
\end{equation}

\noindent The quantity $\nu \sigma_T n_e$ (where $\nu$ is frequency) is Lorentz invariant\cite{Rybicki1986}, so we have $\sigma_T = \sigma_T' / \Gamma^2$, where $\sigma_T'$ is the cross section in the rest frame of the gas. Ultimately, we find that the photospheric radius $R_\mathrm{ph}$ (the radius where $\tau=1$) is

\begin{equation}
    R_\mathrm{ph} = \frac{1.1 \times 10^{11}\,\mathrm{cm}}{ \Gamma^2 \beta^3 \eta}\, ,
\end{equation}

\noindent where $\beta=v/c$. Requiring $R_\mathrm{ph}$ to be smaller than the radius inferred from the light-crossing time, we find

\begin{equation}
    \gamma^4 \beta^3 > 0.06\, \eta^{-1}\, .
\end{equation}

\noindent We obtain $\beta \gtrsim 0.4$ for $\eta=1$ and $\beta \gtrsim 0.6$ for $\eta=0.1$. So, the outflow must be fast, but need not be fully relativistic.

Given that LFBOTs with light curves similar to that of \at\ are rare, occurring at $<0.1\%$ of the core-collapse supernova rate\cite{Ho2023}, and only $\sim10$ LFBOTs have been discovered thus far, it is unlikely that the outflow in \at\ is as tightly collimated as the jets in GRBs (for which $\sim1/100$ events are observed on-axis).
In the extreme case that all the ZTF LFBOTs produced similar outflows, and that \at\ was the only member of the class viewed on-axis so far (although flares cannot be ruled out for all but one of the previously discovered LFBOTs; Methods section~\ref{sec:flare-search}), we estimate a beaming fraction of $f_b = 1/6 = 1-\cos{\theta}$, and find $\theta\approx30^{\circ}$ for the opening angle of the outflow. This estimate of the opening angle is consistent with the current (limited) radio limits on off-axis jets in such objects: the radio emission in AT2018cow (by far the most nearby event, with the most sensitive limits) cannot\cite{Margutti2019} rule out an off-axis jet with $\theta=30^{\circ}$ and energy $E_J<10^{51}\,$erg.

\section{Analysis of Early Optical LFBOT Emission}
\label{sec:lfbot-lc-analysis}

The peak-light measurements of \at\ are well described by a blackbody. From the \at\ ZTF+PS1 $gri$ measurements, we infer $T_\mathrm{eff}=(3.3\pm 1.8) \times 10^{3}$\,K and $R_\mathrm{ph}=(6.8\pm3.0)\times10^{14}\,$cm or $45\pm20$\,AU. These values are very close to those of AT2018cow at peak light. We do not have similar constraints on the blackbody parameters during the decline, but we note that the photospheric radius of AT2018cow's optical emission reached $6\times10^{13}\,$cm by 60\,d\cite{Perley2019,Chen2023a} and $10^{12}\,$cm by 700\,d\cite{Chen2023b}.

The fact that the inferred blackbody radius at peak optical light is much larger than the inferred size of the emitting region during the flares could have several possible explanations. One possibility is that during the first month (when no flares were detected), the blackbody-emitting region expanded enough to become optically thin, finally enabling the smaller flare-emitting region to be observed. 
Another possibility is geometric: that the component producing the flares is on-axis, while the optically thick blackbody-emitting region is off-axis.
Finally, it could be that the flares arise from a jet that took time to burrow through the optically thick material, leaving an open passage through which we are observing.

The rapid fade rate of AT2018cow imposed a limit on the nickel mass\cite{Perley2019,Margutti2019} of $M_\mathrm{Ni}<0.1\,M_\odot$. The slower fade rate of AT2020mrf implied\cite{Yao2022} a limit of $M_\mathrm{Ni}\lesssim 0.26\,M_\odot$. The light curve of \at\ is not well sampled on the decline, but as shown in Figure~\ref{fig:opt-mm-xray-lc} is close to being able to accommodate the light curve of SN\,1998bw, which had\cite{Cano2017} a nickel mass of 0.3--0.6\,$M_\odot$. However, the spectrum at close to $\Delta t=30$\,d showed no supernova features, suggesting that the emission is still dominated by another mechanism.

The persistent blue colours of AT2018cow led to the suggestion that the optical light curve could be powered by reprocessing of the central X-ray source\cite{Margutti2019}, while the light curve of AT2020mrf was found to redden over time\cite{Yao2022}. Although the peak colour of \at's light curve is clearly blue, we have limited information on the colour of the underlying light curve during the decline. A NOT observation at $\Delta t=26\,$d shows $g=21.85\pm0.07\,$mag, $r=22.11\pm0.09\,$mag, and $i=21.98\pm0.10\,$mag; however, the observations consisted of only a single exposure in each filter, and the source was known to have started flaring at this time (from the detection of flares with ZTF), so the contribution of variability and flaring to the observed colour is unclear.

\section{Analysis of Radio Emission}
\label{sec:radio-analysis}

For previously observed LFBOTs, the radio emission has been modeled using a standard equipartition analysis, commonly used in the supernova literature\cite{Chevalier1998}. This framework assumes that the peak frequency is the synchrotron self-absorption frequency\cite{Ho2019,Margutti2019,Coppejans2020,Ho2022_AT2020xnd,Bright2022,Yao2022} and that the underlying electron population has been shock-accelerated into a power-law electron energy distribution. The steep above-peak spectral indices observed in several objects (AT2018cow at early times\cite{Ho2019}, CSS161010\cite{Coppejans2020}, and AT2020xnd\cite{Ho2022_AT2020xnd}) has also been used to argue that the underlying electron population may instead be a relativistic Maxwellian\cite{Margalit2021}.

In this section, we apply a similar analysis to the radio emission from \at. At $\Delta t=40\,$d, the spectral index is $f_\nu \propto \nu^{-0.5\pm0.3}$ (\ref{fig:radio}), consistent with expectations for optically thin emission from a power-law distribution of electrons in the slow-cooling regime. We assume a constant fraction of energy in electrons and magnetic fields, i.e., $\epsilon_e=\epsilon_B=1/3$. This gives a shock radius of\cite{Chevalier1998}

\begin{equation}
    R = (8.8 \times 10^{15}\,\mathrm{cm}) \left(\frac{F_p}{\mathrm{Jy}}\right)^{9/19} \left(\frac{D_A}{\mathrm{Mpc}}\right)^{18/19} \left(\frac{\nu_p}{5\,\mathrm{GHz}}\right)^{-1}.
\end{equation}

\noindent At $\Delta t=40\,$d, the peak flux density $F_p\gtrsim0.3\,\mathrm{mJy}$, and the peak frequency $\nu_p\lesssim100\,\mathrm{GHz}$ (both rest frame). The angular diameter distance $D_A=3\times10^{27}\,\mathrm{cm}$. So, we find a shock radius of $R\gtrsim6\times10^{15}\,$cm and an implied mean shock speed until that time of $v\gtrsim0.06c$, among the slowest inferred radio ejecta speeds for LFBOTs, but very similar to AT2020mrf\cite{Yao2022} (\ref{fig:radio}). 

We can estimate the magnetic field strength of the shock as\cite{Chevalier1998}

\begin{equation}
    B = (0.58\,\mathrm{G}) \left(\frac{F_p}{\mathrm{Jy}}\right)^{-2/19} \left(\frac{D_A}{\mathrm{Mpc}}\right)^{-4/19} \left(\frac{\nu_p}{5\,\mathrm{GHz}}\right).
\end{equation}

\noindent We find $B \lesssim 6\,$G. Using the energy in magnetic fields $U_B$, the total shock energy $U$ is\cite{Ho2019}

\begin{equation}
    U = \frac{U_B}{\epsilon_B} = (1.9 \times 10^{46}\,\mathrm{erg}) \frac{1}{\epsilon_B} 
    \left(\frac{F_p}{\mathrm{Jy}}\right)^{23/19} \left(\frac{D_A}{\mathrm{Mpc}}\right)^{46/19} \left(\frac{\nu_p}{5\,\mathrm{GHz}}\right)^{-1}.
\end{equation}

\noindent We find $U\lesssim 3\times10^{48}\,\mathrm{erg}$. Finally, we can estimate the ambient density $n_e$ as\cite{Ho2019}

\begin{equation}
    n_e = (20\,\mathrm{cm}^{-3}) \frac{1}{\epsilon_B} \left(\frac{L_p}{10^{26}\,\mathrm{erg}\,\mathrm{s}^{-1}\,\mathrm{Hz}^{-1}}\right)^{-22/19} 
    \left(\frac{\nu_p}{5\,\mathrm{GHz}}\right)^{4}
    \left(\frac{t_p}{1\,\mathrm{d}}\right)^{2}.
\end{equation}

\noindent Taking $L_p\gtrsim6\times10^{29}\,\mathrm{erg}\,\mathrm{s}^{-1}\,\mathrm{Hz}^{-1}$, we find $n_e\lesssim 6\times10^{5}\,\mathrm{cm}^{-3}$, close to the value inferred for AT2018cow at $\Delta t=22\,$d\cite{Ho2019}.
So, although we infer near-relativistic velocities from the optical flares (Methods section~\ref{sec:flare-origin}), we infer nonrelativistic shock speeds from the radio emission, implying that the radio emission is not always probing the fastest-moving material in LFBOTs. 

\section{Host Galaxy of \at}
\label{sec:hostgalaxy}

We fit the broadband photometry, which we extracted with the software package \texttt{LAMBDAR}\cite{Wright2016} from the Pan-STARRS images\cite{Chambers2016}, and the absolute-flux-calibrated Keck spectrum from \at\ with the software package \texttt{Prospector} version 1.2.1\cite{Johnson2021}. This program uses the \texttt{Flexible Stellar Population Synthesis} (\texttt{FSPS}) code\cite{Conroy2009} to generate the underlying physical model and \texttt{python-fsps}\cite{ForemanMackey2014} to interface with \texttt{FSPS} in \texttt{python}. The \texttt{FSPS} code also accounts for the contribution from the diffuse gas based on \texttt{Cloudy} models\cite{Byler2017}. We use the dynamic nested sampling package \texttt{dynesty}\cite{Speagle2020} to sample the posterior probability.

We note that the wavelength range of the Keck spectrum was limited to $\lambda_{\rm rest}=3525$--6700\,\AA. The lower cutoff is set by the lower bound of the stellar library \texttt{MILES}\cite{SanchezBlazquez2006} used in \texttt{Prospector}. The upper cutoff is set by the data quality of the Keck spectum.

We assume a simple galaxy model: a Chabrier initial-mass function (IMF)\cite{Chabrier2003} and a linearly increasing star-formation history (SFH) at early times followed by an exponential decline at late times (functional form $t \times \exp\left(-t/\tau\right)$, where $t$ is the age of the SFH episode and $\tau$ is the $e$-folding timescale). This model is attenuated with the Calzetti\cite{Calzetti2000} model. 

\ref{fig:host-fit} shows the observed photometry (black data points) and spectrum (grey), and the best fit (blue). The shaded region indicates the region of the spectrum used in the \texttt{Prospector} fit. We measure a mass of the living stars in the host galaxy of $\log(M/M_\odot)=9.96^{+0.06}_{-0.09}$ and a star-formation rate of $0.55^{+1.36}_{-0.19}\,M_\odot\,$yr$^{-1}$.

\section{Progenitor of \at}
\label{sec:at2022tsd-origin}

The fast timescale of the LFBOT, the luminous and variable X-ray emission, the shallow radio SED peaking in the sub-mm bands, and the characteristics of the optical flares (Methods section~\ref{sec:multiwavelength-properties}, Methods section~\ref{sec:flare-origin}) all support the idea that \at\ involves a near-relativistic outflow powered by a compact object for months.
In addition, as with previous LFBOTs such as AT2018cow and AT2020xnd, the X-rays cannot arise from an extension of the synchrotron spectrum from the radio-emitting electrons\cite{Margutti2019,Ho2019,Ho2022_AT2020xnd}: although the spectral index connecting the millimeter to X-ray emission could be consistent with optically thin synchrotron (\ref{fig:full-sed}), the spectral index of the X-ray emission is not consistent.
The X-rays could potentially arise from inverse-Compton scattering of the ultraviolet-optical photons off the radio-emitting electrons; however, we do not have sufficient data to measure the temporal decay index of the optical light curve during the same period of time as the X-rays were observed.

In this section, we discuss the implications of the above properties for the physical origin of \at\ and other LFBOTs. The location of \at\ at $\sim6\,$kpc from the centre of a dwarf star-forming galaxy (Figure~\ref{fig:transient-parameter-space}; Methods section~\ref{sec:hostgalaxy}), and the fast timescale of the LFBOT, strongly disfavour a supermassive black hole as the compact object. So, we consider stellar- and intermediate-mass black hole engines, both of which have been proposed to explain LFBOTs\cite{Perley2019,Margutti2019,Metzger2022,Chen2023b}. 

The first possibility we consider is that \at\ is powered by a stellar-mass compact object. LFBOTs have been argued to arise from failed supernovae\cite{Perley2019,Margutti2019} or alternatively by the merger of a compact object with a star\cite{Metzger2022}. In these scenarios, there could be three possible energy sources: magnetospheric activity, rotational spindown (for a neutron star), or accretion (for a black hole). We strongly disfavour a magnetospheric energy origin: the total radiated energy in X-rays alone exceeds $10^{50}\,$erg, while the energy in each flare is $\sim10^{47}\,$erg, and the magnetic energy budget of a magnetar would be challenging: $U_B = (2\times10^{49}\,$erg)$(B/10^{16}\,\mathrm{G})^2 (R/10\,\mathrm{km})^3$. However, both rotation or accretion could be possible, very similar to what was argued to explain the TDE candidate J1644+57 as a massive-star collapse event\cite{Quataert2012}. 

For a stellar-mass compact object, the luminosity of the X-ray emission and optical flares ($10^{44}\,$erg\,s$^{-1}$) is highly super-Eddington: $L=10^{6}\,L_\mathrm{Edd}\,(M/M_\odot)$. Such a luminosity is compatible with our inference of a near-relativistic outflow or jet (Methods section~\ref{sec:flare-origin}), which could reduce the intrinsic luminosity by several orders of magnitude. As in J1644+57, the jet would have to be powered for 100\,d, which means that for a core collapse followed by black hole accretion scenario\cite{Woosley1993,WoosleyHeger2012,Kashiyama2015}, the progenitor would have to be extended (a red supergiant\cite{Quataert2012}). Therefore, the failed explosion of a rapidly rotating red supergiant is one plausible progenitor. The prolonged high accretion rate would also be compatible with the merger and tidal disruption scenario\cite{Metzger2022}.

A challenge for the stellar-mass compact object scenario is the minute- to hour-timescale of the flares. By analogy to known flaring systems (Table~\ref{tab:flaring-classes}), possible flare mechanisms are shocks\footnote{If the emission is shock-powered, the variability timescale means it would have to arise from internal rather than external shocks: external shocks cannot\cite{Kumar2015} produce bursts with $\delta t\ll T$.
}, magnetic reconnection events, or turbulence in the jet; the flares themselves could also arise from geometry (jet precession, orbital motion in the case of a binary). For most of these physical mechanisms, the flare duration should scale with the black hole mass, and the duration should be related to the light-crossing time of the black hole. 
For example, for Sagittarius A* the time between flares is $10^{3}$--$10^{4}$ times the light-crossing time $t_\mathrm{cross}$, which is $t_\mathrm{cross}=2GM/c^3=10\,\mathrm{s}$ for a $10^6\,M_\odot$ black hole.
So, a supermassive black hole can have time intervals as long as a day; scaling this down to 1--$10^2\,M_\odot$ would give 1--10\,s as the time between flares, which is clearly far too short.
To explain the long flare durations, the source of the variability would have to be far from the compact object, likely in the outer regions of an accretion disk\cite{Metzger2022}. This could also be a reason to favour an accretion source for the energy, rather than rotation.

Another possible explanation for the flare durations is that the central engine is an intermediate-mass black hole (IMBH). An IMBH TDE was found to be consistent with the LFBOT observed in AT2018cow\cite{Kuin2019,Perley2019}, and an accretion disk around an IMBH was found to be a more natural explanation for the long-lived ultraviolet (UV) emission than a stellar-mass black hole\cite{Chen2023b}. The variable X-ray light curve decaying as $t^{-2}$ is similar to what has been observed in relativistic SMBH TDEs. However, the IMBH picture for AT2018cow is challenged\cite{Margutti2019,Metzger2022} by the presence of extended dense circumburst matter\cite{Ho2019,Nayana2021}, and the occurrence of LFBOTs in host-galaxy environments that resemble those of core-collapse supernovae\cite{Lyman2020}. 

Although IMBH TDEs remain a possibility, we consider the simplest explanation for LFBOTs to be massive-star core-collapse events. In this scenario, \at\ involves a near-relativistic outflow powered by accretion onto a stellar-mass compact object, i.e., a very long-duration GRB analog\cite{Quataert2012}, with high angular momentum from the collapse and accretion of an outer envelope in the failed explosion of an extended star\cite{Perley2021,Metzger2022}, or from the merger and tidal disruption of a star by a stellar-mass black hole\cite{Metzger2022}. The accretion disk gives rise to the significant asphericity observed\cite{Maund2023}, and the flares arise from a process occurring far from the compact object, such as in the outer edges of the accretion disk, or where the outflow dissipates its kinetic energy into radiation. The lack of detected flares in AT2018cow (Methods section~\ref{sec:flare-search}) could be due to viewing angle: AT2018cow is thought to have been observed close to the plane of the circumburst ``disk,'' rather than face-on\cite{Margutti2019,Chen2023b}. A different viewing angle for \at\ could also help to explain the significantly more luminous X-ray emission.
If this association is correct, high-cadence follow-up optical observations of future LFBOTs could reveal the beaming angle of their outflows.

\section{Data for Optical Parameter Space of Different Transient Classes}
\label{sec:data-transient-parameter-space}

Figure~\ref{fig:transient-parameter-space} plots \at\ in optical transient parameter space.
We include data for
core-collapse supernovae (CC~SNe\cite{PerleyBTS,Ho2023}), Type~Ia SNe\cite{PerleyBTS}, superluminous SNe (SLSNe\cite{PerleyBTS}), luminous fast blue optical transients (LFBOTs\cite{Prentice2018,Perley2019,Ho2019,Margutti2019,Perley2021,Ho2020_Koala,Yao2022,Coppejans2020,Ho2022_AT2020xnd,Arcavi2016,Pursiainen2018}), long-duration gamma-ray burst (LGRB) afterglows\cite{Racusin2008,Kann2010}, 
a blazar flare\cite{Nesci2021}, the kilonova AT2017gfo\cite{Kasliwal2017,Villar2017,Cowperthwaite2017,Drout2017}, the optically discovered relativistic TDE AT2022cmc\cite{Andreoni2022}, and the first peak in the optical light curves of low-luminosity GRBs\cite{Galama1998,Campana2006,DElia2018,Ho2020_SN2020bvc}.
Measurements are as close as possible to the rest-frame $g$ band.
Light curves to the upper left of the dashed line\cite{Kasen2017} cannot be powered by the decay of radioactive isotopes because the nickel mass $M_\mathrm{Ni}$ would exceed the ejecta mass $M_\mathrm{ej}$. For the LGRB optical flashes, we started with a sample of LGRB afterglows\cite{Kann2010} and kept light curves that had either a well-resolved peak or observations that started within 100\,s of the burst.

To measure the duration of the light curve of \at, we interpolated the light curve and determined the amount of time the transient spent above half-maximum of peak. We performed a Monte Carlo with 500 samples; the measurement plotted is the mean and the error bar is the standard deviation. The error bar on the peak absolute magnitude is the 1$\sigma$ confidence interval. 

\section{Data for Optical, X-ray, and Millimeter Light curves of Different Transient Classes}
\label{sec:data-transient-lc}

% \bibitem{%1H%Y} %5l %T. \emph{%j}, \textbf{%V}, %p (%Y). \n

Figure~\ref{fig:opt-mm-xray-lc} plots optical, millimeter, and X-ray light curves of different extragalactic transients. In the optical panel, the LFBOT data are of AT2018cow\cite{Perley2019,Ho2019,RiveraSandoval2018,Margutti2019}, AT2020xnd\cite{Perley2021,Ho2022_AT2020xnd,Bright2022}, and AT2020mrf\cite{Yao2022}. We show the optical light curve of the stripped-envelope supernova SN\,1998bw\cite{Galama1998} (GRB\,980425).
Light curves of AT2018cow and AT2020xnd have been scaled to the redshift of \at; the light curve of AT2020mrf has been shifted to match the peak luminosity of \at. The millimeter panel shows relativistic TDEs\cite{Zauderer2011,Yuan2016,Andreoni2022}, LGRBs\cite{Sheth2003,Perley2014,Laskar2018,Laskar2019}, low-luminosity GRBs (LLGRBs\cite{Kulkarni1998,Perley2017}), CC~SNe\cite{Weiler2007,Soderberg2010,Horesh2013,Corsi2014,Maeda2021}, and LFBOTs\cite{Ho2019,Ho2022_AT2020xnd}. For clarity, points marking \at\ are outlined. The X-ray panel shows TDEs\cite{Mangano2016,Andreoni2022}, LFBOTs\cite{RiveraSandoval2018,Margutti2019,Ho2019,Coppejans2020,Ho2022_AT2020xnd,Bright2022,Yao2022}, LGRBs\cite{Yao2022}, LLGRBs\cite{Kouveliotou2004,Tiengo2004,Campana2006,Soderberg2006,Margutti2013}, and CC~SNe\cite{Dwarkadas2012}. For clarity, points marking AT2022tsd and AT2020xnd are outlined.

\section{Data for Table of Flaring Sources}
\label{sec:data-table}

Table~\ref{tab:flaring-classes} summarises the properties of high-amplitude ($\gtrsim 10\times$) flares from a variety of source classes,
including the peak luminosity $L_\mathrm{flare}$, the amplitude (ratio of the flare to the persistent flux; Amp.), and (when applicable) how long the flaring lasts after the main transient event.
Classes include ultraluminous X-ray sources (ULXs\cite{Mucciarelli2007}); a mysterious flaring source GRB\,070610 thought to be Galactic in origin\cite{Kasliwal2008,Stefanescu2008,CastroTirado2008};
neutron star (NS) phenomena such as giant flares (GFs) from soft gamma-ray repeaters (SGRs\cite{Svinkin2021,Frederiks2007}) and nanoshots from the Crab pulsar\cite{Hankins2003}; stellar-mass black hole systems such as X-ray binaries (XRBs\cite{Fender1997}) in the Milky Way and GRBs\cite{Racusin2008} in distant galaxies; and supermassive black hole systems including TDEs\cite{vanVelzen2021,Payne2022}, Sagittarius A*\cite{Marrone2008}, M87\cite{Abramowski2012}, blazars\cite{Nesci2021}, and events displaying quasi-periodic eruptions (QPEs\cite{Miniutti2023}).

\section{Data for Radio Parameter Space Plot}
\label{sec:data-radio-parameter-space}

In \ref{fig:radio}, we show a plot that is commonly used to characterise radio transients\cite{Chevalier1998}.
We include data for CC~SNe (Type~II\cite{VanDyk1993,Weiler1986} and Type~Ib/Ic\cite{Soderberg2005,Soderberg2006,Soderberg2010,Salas2013,Corsi2014}), TDEs\cite{Alexander2016}, LLGRBs\cite{Kulkarni1998,Soderberg2006,Margutti2013,Laskar2017}, LFBOTs\cite{Margutti2019,Ho2019,Coppejans2020,Ho2020_Koala,Yao2022,Ho2022_AT2020xnd}, and two objects discovered by radio surveys (RT\cite{Dong2021,Mooley2022}). Lines of constant shock speed ($R/\Delta t$) are shown, as well as lines of constant mass-loss rate $\dot{M}$ (scaled to wind velocity $v$) in units of $10^{-4}\,M_\odot$\,yr$^{-1}/1000\,$km\,s$^{-1}$. The lines assume that the radio peak is due to synchrotron self-absorption\cite{Chevalier1998}. 

\section{Observations and Data Processing}

\subsection{Palomar 48-inch Samuel Oschin Telescope}
\label{sec:p48}

\at\ was discovered in data from the Zwicky Transient Facility (ZTF\cite{Graham2019,Bellm2019}) custom mosaic camera\cite{Dekany2020}, which is mounted on the 48-inch Samuel Oschin Telescope (P48) at Palomar Observatory.
Three custom filters are used ($g_{\mathrm{ZTF}}$, $r_{\mathrm{ZTF}}$, and $i_{\mathrm{ZTF}}$\cite{Dekany2020}),
and images reach a typical dark-time limiting magnitude of $r\approx20.5\,$mag.
ZTF images are processed and reference-subtracted\cite{Zackay2016}
by the IPAC ZTF pipeline\cite{Masci2019}.
Every 5$\sigma$ point-source detection is saved as an ``alert.''
Alerts are distributed in Avro format\cite{Patterson2019} and to discover \at\ were filtered based on a machine-learning ``real-bogus" metric\cite{Duev2019}, a star-galaxy classifier\cite{Tachibana2018}, and light-curve properties.

Point-spread-function (PSF)-fit forced photometry was performed on archived difference images from the ZTF survey using the ZTF forced-photometry service\cite{Masci2019}. The J2000 coordinates supplied to the service were RA, Dec = 50.0453078, 8.7488721 (decimal degrees), the coordinates of \at\ in the first ZTF alert. The date range was 17 March 2018 (the default value for the beginning of the ZTF survey) to 30 December 2022.
Observations obtained $\geq 15$\,d prior to the first ZTF alert for \at\ all originated from the same ZTF field (506), CCD ID (03), and CCD quadrant (03). 

We followed forced-photometry service guidelines\footnote{\url{https://irsa.ipac.caltech.edu/data/ZTF/docs/forcedphot.pdf}} to further process the data.
We verified that the $r$- and $g$-band reference images were constructed using ZTF images from 2018, years prior to the transient. The $i$-band reference image was constructed using ZTF images from as late as 30 September 2022, but since reference images are constructed using outlier-trimmed averaging\cite{Masci2019} this is unlikely to affect our results; the only $i$-band detection was a flare seen in a single image. 
Four of the observations obtained $\geq 15$\,d prior to the first ZTF alert for \at\ were flagged as being possibly impacted by bad pixels (with the \texttt{procstatus==56} warning).
Two of the four images were available via IPAC; visual inspection showed that the bad-pixel region was 8$''$ from the transient position, sufficiently far away to not impact the photometry, so we kept them in our measurements.
The remaining two images were not available, so we removed them to be conservative.
To identify images impacted by bad weather conditions, we examined the \texttt{zpmaginpsci}, \texttt{zpmaginpscirms}, and \texttt{scisigpix} metrics. We identified two images with outlier values of \texttt{zpmaginpsci<25.5} and removed them.
For each filter, we determined the median flux value of all measurements prior to 10\,d before the first ZTF alert of \at. We subtracted this median value from the flux measurements before converting them to magnitudes.
Finally, we ensured that the PSF-fit reduced $\chi^2$ values had an average value of $\sim 1$ for observations in each filter.
A signal-to-noise ratio (S/N) threshold of 3 was used to identify detections. Nondetections are reported as 5$\sigma$.

\subsection{Pan-STARRS}
\label{sec:panstarrs}

We performed forced photometry on images from the Panoramic Survey Telescope and Rapid Response System (Pan-STARRS1\cite{Tonry2012,Chambers2016,Flewelling2020}).
The typical PS1 observing sequence is $4\times45$\,s per night, with the
four exposures separated over 1\,hr. Filters are $i$, $w$, and $z$\cite{Tonry2012}.
We detected two high-significance (6.4$\sigma$ and 7.9$\sigma$) flares (at $\Delta t=71.1$\,d and $\Delta t=81.1\,$d; Figure~\ref{fig:opt-mm-xray-lc}; \ref{fig:flare-collage}).
In addition, the high-cadence observations during the transient event show no variability, supporting the idea that there is an underlying ``LFBOT'' distinct from the optical flares.

\subsection{ATLAS}
\label{sec:atlas}

We obtained forced photometry at the position of \at\ from the Asteroid Terrestrial-impact Last Alert System (ATLAS\cite{Tonry2018,Smith2020,Shingles2021}).
ATLAS surveys the sky in cyan ($c$) and orange ($o$) filters that are similar to the PS1 $g+r$ and $r+i$ filters,
with a 1\,d cadence.
In three $o$-band observations, we have three low-significance (formally $<3\sigma$) detections at the position of \at.
Stacking the observations results in a clear detection, so we consider these reliable flux measurements.

\subsection{Liverpool Telescope}
\label{Methods: LT}

We obtained $g$- and $r$-band images of \at\ using the IO:O camera on the Liverpool Telescope\cite{Steele2004} (LT) on 15 different nights, from 2022 September 23 to 2023 January 23. We performed astrometric alignment on images that had been reduced using the standard LT pipeline.  Image subtraction was conducted using PS1 as a reference and a custom IDL routine (the PS1 image was convolved to match the PSF of the LT image, then subtracted).  Transient photometry was performed using seeing-matched aperture photometry fixed at the transient location, and calibrated relative to a set of SDSS secondary standard stars in the field (as measured from the unsubtracted images). The LT photometry of \at\ is presented in Supplementary Table 1.

\subsection{Thai National Telescope}
\label{sec:ultraspec}

\at\ was observed with ULTRASPEC\cite{Dhillon2014}, a high-speed imaging photometer mounted on the 2.4\,m Thai National Telescope. Each frame had a 30\,s exposure time, with 15\,msec of dead time between frames. The first epoch was on 2022 December 19, and consisted of 406 $r$-band frames, followed by a 2\,min break to adjust the position of the lower telescope dome shutter, and then by another 161 $r$-band frames. The second epoch was on 2022 December 20, and consisted of 387 $g$-band frames, a 2\,min break, then an additional 91 frames.  Images were taken in $2\times2$ binning, leading to a slight undersampling of the PSF (0.9\arcsec\ pixels in $\sim 2$\arcsec\ seeing).  Image subtraction and photometry were performed relative to PS1 using the same methods and codes as the LT analysis, but with a fixed 2\arcsec\ radius aperture.

\subsection{Himalayan Chandra Telescope}
\label{sec:HCT}

We observed \at\ with the 2\,m Himalayan Chandra Telescope (HCT) on 2022 December 26 under a Director's Discretionary Time proposal. We obtained a series of 5\,min exposures in the $R$ band from 13:47 to 20:25, covering almost all of the first {\it Chandra X-ray Observatory} observing window.  Seeing and focus were generally poor and vary greatly over the course of the observation.  A stacked subset of the best-quality images is used as a reference and all other images are differenced relative to this one by cross-convolution of the respective PSFs.   We did not detect any clear flares, with a limiting magnitude per exposure of $R\gtrsim22\,$mag. It is possible that there are some weak flares at the detection threshold, but the detections are not robust owing to the variable PSF size and shape over the course of the observation window.

\subsection{GROWTH India Telescope}
\label{sec:GIT}

We observed \at\ on 26 December 2022 using the GROWTH-India Telescope (GIT\cite{Kumar2022}) located at the Indian Astronomical Observatory (IAO), Hanle-Ladakh, simultaneously with the Himalayan Chandra Telescope (see previous section). Images were observed in an open filter configuration with a 300\,s exposure time. Images were analysed using a method similar to the one employed on other facilities.  We used a stacked image containing all observations from the night as the reference image to subtract host-galaxy emission in the region of the transient, and performed forced aperture photometry using a 2\arcsec\ radius aperture.  No significant flares were detected during the observation sequence.

\subsection{Magellan-Baade Telescope}
\label{sec:magellan}

Starting at 04:30 on 2022 December 15, we obtained five 3\,min $g$-band exposures of \at\ using the Inamori-Magellan Areal Camera \& Spectrograph (IMACS\cite{Dressler2011}) mounted on the 6.5\,m Magellan-Baade telescope at Las Campanas Observatory.  This sequence shows an unambiguous, high-S/N ($\sim 70$) flare detection peaking in the middle of the five-exposure sequence, and is what led to our initial visual discovery of the short-timescale behaviour of this event.  Image subtraction is performed using a stack of flare-free $g$-band images from Keck/LRIS taken in January as a reference, and forced aperture photometry is applied to the difference image.

\subsection{Nordic Optical Telescope}
\label{sec:not}

Starting at 02:30 on 2022 October 4, we obtained an epoch of $ugri$ observations of \at\ using the Alhambra Faint Object Spectrograph and Camera (ALFOSC) on the 2.56\,m Nordic Optical Telescope (NOT) at the Observatorio del Roque de los Muchachos on La Palma (Spain). Following the discovery of flaring, we obtained two additional epochs of observations, the first in $g$ (five 60\,s exposures the night of 2022 December 16) and the second in $g$ and $r$ ($5 \times 90$\,s exposures in each) the night of 2022 December 23. A flare was detected in the final $g$-band epoch.  Image subtraction in $g$ is performed using a stack of the 2022-12-16 epoch as a reference; image subtraction in $r$ is performed using a stack of the 2022-12-22 observations.  Individual flare-free exposures from the Keck/LRIS observations are used as references for $i$ and $u$.  Photometry is performed using a fixed aperture of 1\arcsec\ radius.
The NOT photometry is presented in Supplementary Table 1.

\subsection{Palomar Hale 200-inch}
\label{sec:chimera}

On 2023 January 27, we observed the position of \at\ for 3\,hr using the Caltech HIgh-speed Multi-color camERA (CHIMERA\cite{Harding2016}) on the Palomar 200-inch Hale telescope. The seeing was 2.5--3\arcsec.
A total of 210 exposures of 50\,s each were obtained simultaneously in the $g$ and $r$ filters. Images were reduced using a custom pipeline modified from that of ULTRACAM\cite{Dhillon2007}, and image subtraction was performed using PS1 as a reference using the same techniques as for LT and ULTRASPEC. Photometry was performed using a 2.5\arcsec-radius aperture.

\subsection{Lulin Observatory}
\label{sec:lulin}

Between 14:38 and 17:27 on 2022 December 26, we obtained 
27 $g$-band images with the Lulin One-meter Telescope (LOT) and 31 $r$-band images with the 40\,cm Super Light Telescope (SLT), coordinated with {\it Chandra X-ray Observatory} observations (Section~\ref{sec:chandra}). Each exposure was 300\,s, with varying seeing conditions (with an average of 2.8\arcsec). The $g$ images were subtracted from a PanSTARRS template, with no detection of \at\ in any image.
Combining all 27 $g$ images results in a 3$\sigma$ limit of $g>22.0$\,mag.
To perform image subtraction on the $r$-band images, a template image was acquired with the SLT.
The 3$\sigma$ upper limits for individual frames are provided in Supplementary Table 1.

\subsection{European Southern Observatory New Technology Telescope}
\label{sec:ntt}

We observed \at\ on two nights (2022 December 18, 19) using ULTRACAM\cite{Dhillon2007}. On December 18 we obtained 116 $i$-band frames with a 20\,s exposure time, totaling 38\,min of data; the deadtime between each frame is 24\,ms. The seeing was 1--1.5\arcsec. On December 19 we obtained 556 $r$-band frames with a 20\,s exposure time, totaling 3\,hr 5\,min of data. The deadtime between each frame is again $\sim 24$\,ms. The seeing started out at 1\arcsec, but worsened to 2.5\arcsec\ toward the end of the run. We subtracted a %(flat-fielded) 
dark image and removed remaining bad/hot pixels in the vicinity of the transient by taking the median value of the eight surrounding pixels.  Image subtraction was performed %in the $g$, $r$, and $i$ filters 
using a consistent method as for the other observations, using stacks formed from flare-free sections of the data taken the same night.  For the first night, which shows no flaring, we use a stack of the entire night; for the second night we use a stack of the first 97 images (all acquired prior to the flare).  %For the $u$-band observations, we did not perform image subtraction as the host galaxy is not detected in a stack, although we did perform a scalar offset to the flux of all exposures using the median flux of the entire sequence.  
Photometry was performed using a fixed 1.5\arcsec-radius aperture and calibrated to nearby Pan-STARRS standards.

As part of ePESSTO+ (the Public European Southern Observatory Spectroscopic Survey of Transient Objects project\cite{Smartt2015}), we observed \at\ on three nights (2022 December 22, 24, and 30) in the $g$ and $r$ bands using the Faint Object Spectrograph and Camera (v.2; EFOSC2\cite{Buzzoni1984}) mounted on the 3.58\,m European Southern Observatory (ESO) New Technology Telescope (NTT) under the observing program 1108.D-0740 (PI C. Inserra). On the first two nights, the observation sequence was $5\times$95\,s exposures in $g$ followed by $5\times95$\,s exposures in $r$.  A flare is seen at the beginning of the $g$-band sequence from the second epoch; otherwise no variability was evident. On the third night, the sequence was altered such that images were obtained in alternating filters ($5\times gr$) and no flare was detected. The data were reduced using the standard pipeline\footnote{\url{ https://github.com/svalenti/pessto}}, which is based on iraf/pyraf. Image subtraction was performed using the last exposure of each sequence as a reference image; photometry was performed using a 1.0\arcsec-radius aperture in all observations.

\subsection{Kitt Peak 84-inch Telescope}
\label{sec:kp84}

On 2022 December 20, we observed the position of \at\ for 2\,hr using the Spectral Energy Distribution Machine (SEDM\cite{Blagorodnova2018}) version 2 on the Kitt Peak 84-inch (KP84) Telescope. A total of 60 exposures of 120\,s each were obtained in the clear filter. Flat-fielding was performed using a super-sky flat constructed using a median stack of all exposures taken on the field.  Pan-STARRS $r$-band imaging was used as the reference image, which resulted in an acceptable removal of the host despite the unfiltered nature of the observations.  Photometry was performed using a fixed 1.5\arcsec-radius aperture and calibrated to nearby Pan-STARRS standards. We subtracted a median flux level from all flux values.

\subsection{Large Array Survey Telescope}
\label{sec:last}

We observed \at\ using eight telescopes in the Large Array Survey Telescope (LAST\cite{Ofek2023,BenAmi2023}).
The target was observed on 2023 January 12, 13, and 15, and also on several nights during December 2022.
The 2022 observations were taken under poor conditions and are not reported here.
We obtained 20\,s exposures in continuous mode (i.e., no dead time between images).
A total of 10.9\,hr of observations in 3 nights were obtained.
The observations were reduced using the LAST pipeline \cite{Ofek2014,Ofek2019,Ofek2023}), and forced PSF photometry was conducted on the individual images
in the transient position. 
The source position was fitted but it was forced to be within 0.5 pixels ($0.62''$) of the initial position.
In each image, we also performed forced photometry on all {\it Gaia}-DR3\cite{Gaia2021} stars
within $500''$ from the transient position.
These sources were used for the photometric calibration.

Since in many cases, we observed the transient location simultaneously with several LAST telescopes, in Supplementary Table 2 we provide a 2\,min binning of the unsubtracted measurements. We did not detect any flares.

\subsection{W.~M. Keck Observatory}
\label{sec:keck}

We obtained five epochs of observations of \at\ using the Low Resolution Imaging Spectrometer (LRIS\cite{Oke1995}) at the W.~M.~Keck Observatory; it is equipped with an atmospheric dispersion corrector. The first epoch, obtained as part of a program with PI A. V. Filippenko, was a 40~min exposure starting at 13:52:48.69 on 2022 September 23. The setup was a $1''$ slit, blue grism 600/4000, red grating 400/8500, and dichroic 560. Binning was $1 \times 1$ in both the red and blue CCDs, and the position angle of the slit was 30$^\circ$ counterclockwise from north. The wavelength coverage was 3138--10,259\,\AA. 

The second epoch was a 40\,min exposure starting at 14:13:16 on 2022 October 6. The setup was a $1''$ slit, blue grism 400/3400, red grating 400/8500, and dichroic 560. Binning was $1 \times 2$ (spatial, spectral) in the blue CCD and $1\times1$ in the red, and the position angle of the slit was 61$^\circ$ counterclockwise from north. The wavelength coverage was 3109--9646\,\AA. The data were obtained as part of a ToO program with PI R. Margutti. 

We obtained two imaging epochs in the $g$ and $I$ bands (PI M. Kasliwal), each comprising four exposures totaling 20\,min. The first epoch started at 2022 October 19 10:35 and the second epoch started at 2023 January 17 07:12. 
Finally, we obtained one imaging epoch in the $u$ and $I$ bands (PI J. Cooke). The observation comprised five exposures of 5~min each, beginning at 10:36 on 2022 December 29.

All spectra and images were reduced using LPipe\cite{PerleyLPipe}.
For the last two image sequences (in December and January), we performed image subtraction using the last image of the sequence as the reference; for the first (October) imaging sequence we use stacks of the January observations as a reference.  Photometry was performed using a 1.25\arcsec-radius aperture.  The $g$ and $I$ images are calibrated relative to PS1.  The $u$-band image was calibrated relative to a LT-IO:O calibration of the field taken on two photometric nights in January 2023.

The pipeline-reduced LRIS spectra show a slight inconsistency between the wavelength calibrations in the blue region owing to flexure, which was rectified using an additional 2\,\AA\ shift calculated using the position of a weak 5200\,\AA\ night-sky line.
Even after this correction, there remains an offset of 2\,\AA\ between host emission-line features in the two Keck spectra, which is apparent in all the lines. The night-sky-line positions are consistent, however, so this is likely due to slightly different slit positions and orientations. 
%Dan: we probably don't really need to discuss this here?

%The line that "shifted" was on the blue, right?   It's much more difficult to exactly wavelength-calibrate the blue CCD versus the red due to the lack of strong night sky lines (also fewer arc calibration lines).  The "edges" of both CCDs also have this problem.    Whereas for the redshift measurement you can choose your lines to be in whatever part of the spectrum is calibrated best (usually the middle of the red side).  So the fact that we have some uncertainty in the blue for particular lines doesn't necessarily mean the redshift is inaccurate.

\subsection{Upgraded Giant Metrewave Radio Telescope}
\label{sec:GMRT}

We triggered upgraded Giant Metrewave Radio Telescope (uGMRT) observations of AT2022tsd
during 2023 March 04.51 to 2023 April 02.42 in frequency bands 1000--1460\,MHz
(Band 5), 550--750\,MHz (Band 4), and 250--500\,MHz (Band 3). The data were recorded
in total intensity mode with bandwidths 400\,MHz (Band 5) and 200\,MHz (Band 4 and Band 3)
split into 2048 channels. The temporal resolution was 10\,s. We used 3C147 as the flux density
calibrator and J0323+055 as the phase calibrator.
The data were analysed\cite{Nayana2017} using the Astronomical Image Processing Software (AIPS\cite{Greisen2003})
The data were initially flagged and calibrated using standard tasks in AIPS. The fully calibrated data were imaged
using task IMAGR. A few rounds of phase-only self-calibration were performed to improve the
image quality. The details of the GMRT observations are presented in \ref{tab:radio-observations}. The quoted errors
include map root-mean-square (RMS) and a 10\% calibration error added in quadrature.

\subsection{Very Large Array}
\label{sec:VLA}

Seven epochs of Karl G.\ Jansky Very Large Array (VLA\cite{Perley2011}) observations were obtained of \at\ from 2022 October 2 to 2023 April 5 under Program ID 2022B-157 and ToO Program ID 2023A-393.
The first epoch was obtained during the D-to-C configuration change, the next four epochs were obtained in the C configuration, and the final two epochs were obtained in the B configuration.
All observations used 3-bit samplers, full polarization, and employed 3C147 and J0321+1221 as flux-density and phase calibrators, respectively.

Data were calibrated using the VLA pipeline available in the Common Astronomy Software Applications (CASA\cite{McMullin2007}).
Epoch 2 was hampered by poor phase stability at high frequencies, affecting the Ka and Q-band observations. Additional flagging was performed manually and the calibration pipeline was rerun, albeit with continued high RMS noise at these high frequencies.
Prior to imaging each observation, additional radio-frequency interference (RFI) was removed by flagging amplitudes higher than $3\sigma$.
For the Epoch 4 Ku-band observation we flagged additional spectral windows manually to excise RFI.

For imaging, we adopted Briggs weighting (\texttt{robust=0.5}) and \texttt{nterms=2}. For some high-frequency observations we adopted natural weighting because it significantly improved the S/N of the image.
The pixel scale was chosen to oversample the beam size by a factor of $\geq10$ in all images.
In each image, we verified that the source was unresolved using \texttt{imfit}.
For the Epoch 4 Ku-band observation the source appeared slightly resolved, perhaps due to underlying diffuse host-galaxy emission, or the fact that the source lies along a sidelobe.
In all cases we adopted the maximum pixel flux as the flux density.
To determine the uncertainty in the flux density we measured the RMS noise in a nearby region of the image unaffected by any sources.

To search for short-timescale variability, we imaged each scan of the 15\,GHz observations individually. We chose 15\,GHz because the VLA is more sensitive at this frequency than at higher frequencies, and because the length of the cycle time is well suited to searching for variability on the timescale of the observed \at\ flares. Each observation had 6--8 scans, each scan lasted $\sim 7$\,min, and scans were typically separated by 1\,min. The resulting S/N per scan ranged from $<3\sigma$ (no detection, most common in Epoch 1 and Epoch 2) to S/N = 8 (in Epochs 5 and 6).
We did not detect any definitive variability. The strongest variations we measured were during Epoch 3 (when the source apparently brightened from $28\pm8\,\mu$Jy to $45\pm8\,\mu$Jy, then faded to nondetection with RMS $8\,\mu$Jy) and Epoch 6 (when the source apparently faded from $70\pm9\,\mu$Jy to $37\pm8\,\mu$Jy across two scans). However, these variations are fairly marginal; in the Epoch 6 observation, the corresponding flux density of another source in the field was $65\,\mu$Jy and then $75\,\mu$Jy, suggesting that the true uncertainty is $\sim 10\,$$\mu$Jy. In that case, the fading is only $\sim 3\sigma$.

Using the B-configuration Ku-band observation, we obtain the following measurement of the position of \at:
standard equinox J2000 right ascension $\alpha$ =
$03^{\rm{h}}20^{\rm{m}}10^{\rm{s}}.873$
and declination $\delta$ = $+08^{\circ} 44' 55''.739$ 
(uncertainty $0.009''$).

\subsection{Submillimeter Array}
\label{Methods:SMA}

\at\ was observed with the SMA on 2022 October 4 with 7 antennas for a total of 5.95\,hr on source, under ToO program 2022A-S019. The atmospheric opacity was poor and variable, changing from 0.28 to 0.18 over the night. Observations were performed using R$\times$A and R$\times$B receivers both tuned to LO frequencies of 225.55\,GHz. All 48\,GHz of bandwidth were used to generate a single continuum channel. Observations of the nearby quasars 0238+166 and 0423-013 were used as the primary phase and amplitude gain calibrators with absolute flux calibration performed by comparison to Neptune and Uranus while passband calibration was derived using BL Lac. Calibration was performed using the MIR IDL package for the SMA, with subsequent analysis performed in MIRIAD. No source was detected. The final image has an RMS of 0.27\,mJy and synthesised beam of 3.9\arcsec\ $\times$ 3.2\arcsec.

\subsection{Atacama Large Millimeter/submillimeter Array}
\label{sec:ALMA}

\at\ was observed with ALMA as part of DD time (Project code 2022.A.00010.T) during Cycle 9 using Bands 6--8. Observations were performed on 2022 October 19 ($\Delta t \approx 43$\,d; Band 7), 2022 October 21 ($\Delta t \approx 45$\,d; Band 8), and 2022 October 22 ($\Delta t \approx 46$\,d; Band 6) with $\Delta t$ epochs in the observer frame. The ALMA 12\,m antenna array was in its C-3 configuration, with
43--46 working antennas and baselines in the range 15.1--457.3\,m. The on-source integration time was
11\,min in Band 6, 50\,min in Band 7, and 2.0\,hr in Band 8.
Observations used dual-sideband (2SB) receivers with a total bandwidth of 7.5\,GHz. The total bandwidth was divided into four 1.875\,GHz basebands centred on 224, 226, 240, and 242\,GHz (Band 6);
336.5, 338.5, 348.5, and 350.5\,GHz (Band 7);
and 398, 400, 410, 412\,GHz (Band 8).

%10m38s in Band 6, 49m42s in Band 7, and 1.99\,hr in Band 8.

All calibration and imaging was done with CASA.
The data were calibrated and imaged
with the standard ALMA pipeline, using J0309+1029 to calibrate the complex gains, and using J0238+1636 (Bands 6 and 7) or J0423-0120 (Band 8) to calibrate the bandpass response and apply an absolute flux scale.
\at\ is unresolved in the Band 6 and Band 7 data, and partially resolved in the Band 8 data (i.e., the fitted width is larger than the synthesised beam).
The S/N in the resulting images is 11 in Band 6, 12 in Band 7, and 7 in Band 8.
The ALMA results are summarised in \ref{tab:radio-observations}.

We searched for variability across each observation. The Band 6 observations started at 04:02 and ended at 04:13 on 2022-10-22, spanning 11\,min. We imaged each of the two scans individually, for a per-scan S/N of 6--9, with no significant difference in the flux density between scans.
The Band 7 observations started at 04:29 and ended at 05:43 on 2022-10-19, spanning 1\,hr 14\,min.
We imaged each of the eight on-target scans individually, for a per-scan S/N of 4--7, and did not detect any significant changes between scans. The time per scan was 4.5--7\,min.
Finally, the Band 8 observations started at 04:55 and ended at 08:10 on 2022-10-21, spanining 3\,hr 15\,min. We imaged each of the 19 on-target scans individually, and did not detect emission from \at\ in any scan.

\subsection{NOrthern Extended Millimeter Array (NOEMA)}
\label{Methods:NOEMA}

We obtained six epochs of observations of \at\ with NOEMA. Multiband observations were done when the source flux and weather permitted it, with Band 1 (100\,GHz), Band 2 (150\,GHz), and Band 3 (230\,GHz) under the ToO program S22BD. A total of 14 observations were obtained, and interferometer array configurations ranged from compact (D) to more extended (C) and (B). The primary flux calibrators were MWC349 and LKHA101, and the time-dependent phase and amplitude calibrators were the QSOs B0306+101 and B0256+075. The data reduction was done with the CLIC software (GILDAS package\cite{GILDAS}). 
Dual-polarization UV tables were written for each of the receiver sidebands. The resulting calibrated UV tables were analysed in the MAPPING software (also from the GILDAS package) and point-source UV plane fits were performed.
The NOEMA results are summarised in \ref{tab:radio-observations}.

We searched for flux variability over the course of the two highest-S/N observations: the Band 2 observation during the night of 2022 October 29--30, and the Band 1 observation during the night of 2022 November 18--19.
The UV point position for the combined data was fit separately for the LSB and the USB, in
order to account for minor calibration errors.
Then, point-source fits were performed to each of the five on-target scans. Each scan lasted 22.5\,min, and the total observation window was 2.5\,hr. The S/N in each scan ranged from 3--4. No significant variability was detected.

\subsection{Neil Gehrels Swift Observatory}
\label{sec:swift}

\at\ was observed by the X-ray Telescope (XRT\cite{Burrows2005}) onboard the {\it Neil Gehrels Swift Observatory} under a series of time-of-opportunity (ToO) requests, with a total of 14 segments. The first segment began at 09:13 on 2022 October 4 ($\Delta t=28.2\,$d, observer frame), and the last segment ended at 21:10 on 2022 December 17 ($\Delta t=102.7\,$d, observer frame). The source was not detected in the last segment, so we did not pursue further XRT observations.
All XRT observations were obtained in the photon-counting mode, and are summarised in \ref{tab:xray-observations}.
The transient was also observed by the Ultra-Violet/Optical Telescope (UVOT\cite{Roming2005}), but the only emission detected was from the host galaxy. 

To measure the count rate from each observation, we used the analysis tools developed by the {\it Swift} team\cite{Evans2007,Evans2009}.
We used iterative centroiding and binned by observation. 
To convert from count rate to unabsorbed flux, we fit for an average spectrum using the first five observations. Using a Galactic neutral hydrogen column density\cite{Willingale2013} of $n_H=2.11\times10^{21}\,$cm$^{-2}$, the data were well described by a power law with photon index $\Gamma=2.1^{+0.5}_{-0.4}$, 
giving a 0.3--10\,keV count rate to flux conversion factor of $5.10\times 10^{-11}\,\rm erg\,cm^{-2}\,ct^{-1}$.
An independent analysis of the {\it Swift} data\cite{Matthews2023} found a consistent value for the photon index of $\Gamma=2.00^{0.17}_{-0.15}$.

\subsection{Chandra X-ray Observatory}
\label{sec:chandra}

\at\ was observed by the \emph{Chandra X-ray Observatory} under two programs (Proposal 24500280, PI D. Matthews; DDT Proposal 23508884, PI A. Ho) for a total of eight epochs. The first epoch began on 2022 October 16 and the most recent epoch began on 2023 July 16. Exposure times ranged from 12\,ks to 40\,ks.
After the detection of the Magellan/IMACS flare, we were granted
40\,ks of {\it Chandra} observations under Director's Discretionary Time, divided into two windows (2022 December 26 and 29), to search for simultaneous X-ray and optical flares.
We conducted simultaneous ground-based optical observations with the Himalayan Chandra Telescope, the Lulin Observatory, and Keck/LRIS (Methods section~\ref{sec:keck}).
A single optical flare was detected with Keck/LRIS on 29 December (\ref{fig:flare-collage}), but no X-ray flare counterpart was detected.

We reduced each epoch using the Chandra Interactive Analysis of Observations (CIAO\cite{Fruscione2006}) software package (v4.15). Counts were extracted from \at\ using a circle with radius $2''$, and background counts were measured in source-free regions near \at.
We used \texttt{specextract} to bin the spectrum (with 5 counts per bin for all epochs). The routine \texttt{sherpa} was used to fit the spectrum in the range 0.5--6\,keV, with the background subtracted, using a model with photoelectric absorption and a single-component power law (\texttt{xsphabs.abs1 $\times$ powlaw1d.p1}). We set the Galactic hydrogen density to be the same as for the \emph{Swift} observations. In all epochs, the data were well described by a power law (reduced $\chi^2=0.2$--1.2). In the highest-S/N observation, we found $\Gamma=1.98\pm0.23$; all other epochs had a best-fit $\Gamma$ consistent with this value.
An independent analysis of the {\it Chandra} data\cite{Matthews2023} found a consistent value for the photon index of $\Gamma=1.89^{+0.09}_{-0.08}$.
After obtaining the best-fit model of the spectrum, we used \texttt{sample\_flux} to measure the 0.5--6\,keV flux of the source. The best-fit flux measurements are listed in Table~\ref{tab:chandra}. To convert to the \emph{Swift} 0.3--10\,keV range (\ref{fig:xray-lc}) we multiplied the 0.5--6\,keV values by a factor of 1.77.

For the final epoch of observations, we binned three observations that were obtained on three different days, close together in time (2023 June 11--16), after confirming by analysing each observation individually that there was no strong variability between epochs. To bin, we used \texttt{merge\_obs} to create a merged file, and used \texttt{srcflux} to compute the count rate. There were insufficient counts to perform a spectral fit, so we adopted the same spectral index as for the other epochs ($\Gamma=2$). 

For each sufficiently bright observation, we used \texttt{dmextract} and 500\,s bins to construct a light curve of \at. We also extracted the light curve of the background region. The light curves of \at\ and the background are shown in \ref{fig:xray-lc}, with 1$\sigma$ error bars.

\noindent{\bf References}
%\bibliographystyleNew{naturemag}
%\bibliographyNew{references}
%\bibliographystyle{naturemag}
%\bibliography{references, gcns}

\end{methods}

\begin{addendum}

\item It is a pleasure to thank the two anonymous referees for their feedback, which greatly improved the content and clarity of the paper.

A.Y.Q.H. would like to thank Eliot Quataert, Dong Lai, Jim Cordes, and Sterl Phinney for discussions on the physical origin of \at\ and its flares; Shami Chatterjee and Dillon Dong for advice on VLA calibration and imaging; Brad Cenko for assistance with {\it Swift} observations; Murray Brightman and Brad Cenko for assistance with {\it Chandra} data reduction; Kimberly Ward-Duong, Kate Follette, Sarah Betti, Jada Louison, Jingyi Zhang, Raffaella Margutti, and Ryan Chornock for assistance with Keck target-of-opportunity (ToO) observations; Ilsang Yoon for advice on ALMA calibration and imaging; and Adam Miller for discussions about optical time-series analysis.
P. C. would like to thank Yuri Beletsky for his assistance in remote observations with the Magellan telescope.
S. Schulze acknowledges support from the G.R.E.A.T. research environment, funded by {\em Vetenskapsr\aa det},  the Swedish Research Council, project number 2016-06012.
VSD, ULTRASPEC, and ULTRACAM are funded by the UK’s Science and Technology Facilities Council (STFC), grant ST/V000853/1.
S.J.S. acknowledges funding from STFC grants ST/T000198/1 and ST/S006109/1.
This work was funded by ANID, Millennium Science Initiative, ICN12\_009.
We thank Lulin staff H.-Y. Hsiao, C.-S. Lin, W.-J. Hou, H.-C. Lin, and J.-K. Guo for observations and data management. 
M.W.C. acknowledges support from the U.S. National Science Foundation (NSF) with grants PHY-2010970 and OAC-2117997. 
L.G., C.P.G, and T.E.M.B. acknowledge financial support from the Spanish Ministerio de Ciencia e Innovaci\'on (MCIN), the Agencia Estatal de Investigaci\'on (AEI) 10.13039/501100011033, the European Social Fund (ESF) ``Investing in your future,” the European Union Next Generation EU/PRTR funds, the Horizon 2020 Research and Innovation Programme of the European Union, and by the Secretary of Universities and Research (Government of Catalonia), under the PID2020-115253GA-I00 HOSTFLOWS project, the 2019 Ram\'on y Cajal program RYC2019-027683-I, the 2021 Juan de la Cierva program FJC2021-047124-I, the Marie Skłodowska-Curie and the Beatriu de Pin\'os 2021 BP 00168 programme, and from Centro Superior de Investigaciones Cient\'ificas (CSIC) under the PIE project 20215AT016, and the program Unidad de Excelencia Mar\'ia de Maeztu CEX2020-001058-M.
A.G.-Y.’s research is supported by the EU via ERC grant 725161, the ISF GW excellence center, an IMOS space infrastructure grant, a GIF grant, as well as the André Deloro Institute for Advanced Research in Space and Optics, The Helen Kimmel Center for Planetary Science, the Schwartz/Reisman Collaborative Science Program, and the Norman E. Alexander Family M Foundation ULTRASAT Data Center Fund, Minerva and Yeda-Sela;  A.G.-Y. is the incumbent of the The Arlyn Imberman Professorial Chair.
Nayana A.J. would like to acknowledge DST-INSPIRE Faculty Fellowship (IFA20-PH-259) for supporting this research
C.C.N. is grateful for funding from the Ministry of Science and Technology (Taiwan) under contract 109-2112-M-008-014-MY3.
M.N. is supported by the European Research Council (ERC) under the European Union’s Horizon 2020 research and innovation programme (grant agreement No.~948381) and by funding from the UK Space Agency.
E.O.O. acknowledges grants from the ISF, IMOS, and BSF.
F.O. acknowledges support from MIUR, PRIN 2017 (grant 20179ZF5KS) ``The new frontier of the Multi-Messenger Astrophysics: follow-up of electromagnetic transient counterparts of gravitational wave sources.''
D.P. is grateful to the LAST Observatory staff.
M.P. is supported by a research grant (19054) from VILLUM FONDEN.
A.V.F.'s group at U.C. Berkeley received financial support from the Christopher R. Redlich Fund, Gary and Cynthia Bengier, Alan Eustace, Sanford Robertson, Clark and Sharon Winslow, Briggs and Kathleen Wood, and many other donors.
The work of D.S. was carried out in the framework of the basic funding
program of the Ioffe Institute No. 0040-2019-0025.

%We are grateful for the expert assistance of the staffs at the many observatories where data were obtained; these are listed below. 

Based in part on observations obtained with the Samuel Oschin Telescope 48-inch and the 60-inch Telescope at Palomar Observatory as part of the Zwicky Transient Facility project. ZTF is supported by NSF grants AST-1440341 and AST-2034437 and a collaboration including current partners Caltech, IPAC, the Weizmann Institute of Science, the Oskar Klein Center at Stockholm University, the University of Maryland, Deutsches Elektronen-Synchrotron and Humboldt University, the TANGO Consortium of Taiwan, the University of Wisconsin at Milwaukee, Trinity College Dublin, Lawrence Livermore National Laboratories, IN2P3, University of Warwick, Ruhr University Bochum, Northwestern University and former partners the University of Washington, Los Alamos National Laboratories, and Lawrence Berkeley National Laboratories. Operations are conducted by COO, IPAC, and UW.
The ZTF forced-photometry service was funded under Heising-Simons Foundation grant \#12540303 (PI: M. Graham).

The Pan-STARRS1 Surveys (PS1) and the PS1 public science archive have been made possible through contributions by the Institute for Astronomy, the University of Hawaii, the Pan-STARRS Project Office, the Max-Planck Society and its participating institutes, the Max Planck Institute for Astronomy, Heidelberg and the Max Planck Institute for Extraterrestrial Physics, Garching, The Johns Hopkins University, Durham University, the University of Edinburgh, the Queen's University Belfast, the Harvard-Smithsonian Center for Astrophysics, the Las Cumbres Observatory Global Telescope Network Incorporated, the National Central University of Taiwan, the Space Telescope Science Institute, the National Aeronautics and Space Administration (NASA) under grant NNX08AR22G issued through the Planetary Science Division of the NASA Science Mission Directorate, NSF grant AST-1238877, the University of Maryland, Eotvos Lorand University (ELTE), the Los Alamos National Laboratory, and the Gordon and Betty Moore Foundation.

This work has made use of data from the Asteroid Terrestrial-impact Last Alert System (ATLAS) project. The Asteroid Terrestrial-impact Last Alert System (ATLAS) project is primarily funded to search for near-Earth objects (NEOs) through NASA grants NN12AR55G, 80NSSC18K0284, and 80NSSC18K1575; byproducts of the NEO search include images and catalogues from the survey area. This work was partially funded by Kepler/K2 grant J1944/80NSSC19K0112 and HST GO-15889, and STFC grants ST/T000198/1 and ST/S006109/1. The ATLAS science products have been made possible through the contributions of the University of Hawaii Institute for Astronomy, the Queen’s University Belfast, the Space Telescope Science Institute (STScI), the South African Astronomical Observatory, and The Millennium Institute of Astrophysics (MAS), Chile.

The Liverpool Telescope is operated on the island of La Palma by Liverpool John Moores University in the Spanish Observatorio del Roque de los Muchachos of the Instituto de Astrofisica de Canarias with financial support from the UK Science and Technology Facilities Council.
Based in part on observations made with ULTRASPEC at the Thai National Observatory, which is operated by the National Astronomical Research Institute of Thailand (Public Organization).
Based in part on observations obtained with the Spectral Energy Distribution
Machine on the Kitt Peak 84-inch telescope (SEDM-KP). The SEDM-KP team
thanks the NSF and the National
Optical-Infrared Astronomy Research Laboratory for making the Kitt
Peak 2.1\,m telescope available. SEDM-KP is supported by the Heising
Simons Foundation under grant 2021-2612 titled ``The SEDM Kitt
Peak Project,"
and a collaboration including current partners Caltech, University of
Minnesota, the
University of Maryland, Northwestern University, and STScI.

This work made use of data from the GROWTH-India Telescope (GIT) set up by the Indian Institute of Astrophysics (IIA) and the Indian Institute of Technology Bombay (IITB). It is located at the Indian Astronomical Observatory (Hanle), operated by IIA. We acknowledge funding by the IITB alumni batch of 1994, which partially supports operations of the telescope. Telescope technical details are available online.\cite{growth_india}
We thank the staff of IAO, Hanle and CREST, Hosakote, that made these obervations possible. The facilities at IAO and CREST are operated by the Indian Institute of Astrophysics, Bangalore.
This paper includes data gathered with the 6.5~m Magellan Telescopes located at Las Campanas Observatory, Chile.
Based on observations made with the Nordic Optical Telescope, owned in collaboration by the University of Turku and Aarhus University, and operated jointly by Aarhus University, the University of Turku, and the University of Oslo, representing Denmark, Finland, and Norway (respectively), the University of Iceland, and Stockholm University at the Observatorio del Roque de los Muchachos, La Palma, Spain, of the Instituto de Astrofisica de Canarias.
This publication has made use of data collected at Lulin Observatory, partly supported by MoST grant 108-2112-M-008-001.
Based partially on observations collected at the European Organisation for Astronomical Research in the Southern Hemisphere, Chile, as part of ePESSTO+ (the advanced Public ESO Spectroscopic Survey for Transient Objects Survey).
ePESSTO+ observations were obtained under ESO program 108.220C (PI: Inserra).

% The KPED team thanks the National Science Foundation and the National Optical Astronomical Observatory for making the Kitt Peak 2.1-m telescope available. The KPED team thanks the National Science Foundation, the National Optical Astronomical Observatory and the Murty family for support in the building and operation of KPED. In addition, they thank the CHIMERA project for use of the Electron Multiplying CCD (EMCCD).

Some of the data presented herein were obtained at the W. M. Keck Observatory, which is operated as a scientific partnership among the California Institute of Technology, the University of California, and NASA. The Observatory was made possible by the generous financial support of the W. M. Keck Foundation.
The authors wish to recognise and acknowledge the very significant cultural role and reverence that the summit of Maunakea has always had within the indigenous Hawaiian community.  We are most fortunate to have the opportunity to conduct observations from this mountain.

%We thank the staff of the GMRT that made these observations possible. 
GMRT is run by the National Centre for Radio Astrophysics of the Tata Institute of Fundamental Research

The Submillimeter Array is a joint project between the Smithsonian Astrophysical Observatory and the Academia Sinica Institute of Astronomy and Astrophysics and is funded by the Smithsonian Institution and the Academia Sinica.
This paper makes use of the following ALMA data: ADS/JAO.ALMA\#2022.A.00010.T. ALMA is a partnership of ESO (representing its member states), NSF (USA), and NINS (Japan), together with NRC (Canada), MOST and ASIAA (Taiwan), and KASI (Republic of Korea), in cooperation with the Republic of Chile. The Joint ALMA Observatory is operated by ESO, AUI/NRAO, and NAOJ. The National Radio Astronomy Observatory is a facility of the NSF operated under cooperative agreement by Associated Universities, Inc.
Based in part on observations carried out with the IRAM Interferometer NOEMA. IRAM is supported by INSU/CNRS (France), MPG (Germany), and IGN (Spain).

This work made use of data supplied by the UK Swift Science Data Centre at the University of Leicester.
The scientific results reported in this article are based in part on observations made by the {\it Chandra} X-ray Observatory. This research has made use of software provided by the Chandra X-ray Center (CXC) in the application packages CIAO and Sherpa.

\item[Contributions] 
All authors reviewed the manuscript and contributed to the source interpretation.
A.Y.Q.H. identified the source; coordinated the follow-up observations; performed radio, X-ray, and some optical data analysis; performed the source analysis and modeling; and wrote the majority of the manuscript. D.A.P. performed the optical image subtraction and photometry, and contributed significantly to the follow-up campaign, the source analysis, and the manuscript. P.C. observed the source with Magellan and LAST, made the first flare identification, and assisted with follow-up observations and X-ray data analysis. S.S. performed the host-galaxy analysis, assisted with {\it Swift} data analysis, and performed follow-up observations with the NOT. V.D. provided follow-up observations with ULTRACAM and ULTRASPEC. H.K., V.S., A.S., and V.B. performed follow-up observations and image reduction with the GIT and the HCT. M.B. performed NOEMA follow-up observations and data reduction. S.J.S. provided ATLAS and Pan-STARRS photometry. J.P.A., C.P.G., L.G., M.G., C.I., T.E.M., F.O., M.P., P.J.P, P.W., Y.W., and D.Y. are ePESSTO+ builders. A.G.C. and S.B. are GIT builders. S.A. and S. P. enabled NARIT/TNT observations. E.B., R.B., M.C., A.D., M.G., A.M., B.R., R.R., and A.W. are ZTF builders. S.B., D.P., E.S., and N.S. are LAST builders. T.B. M.F., H.G., E.M., M.N., K.S., and S.S. assisted with Pan-STARRS data analysis. T.G.B., J.C., C.F., J.F., S.R.K., M.M.K., V.K., and M.S. assisted with Keck follow-up observations. N.A.J. and P.C. performed uGMRT follow-up observations and data analysis. T.-W.C., W.-P.C., C.-C.N., Y.-C.P., and S.Y. assisted with Lulin observations and data reduction. K.K.D. observed with the P200. A.V.F. assisted with Keck follow-up observations and thoroughly reviewed the manuscript. A.G.-Y. assisted with LAST follow-up observations. K.H. assisted with LT follow-up observations. T.M. contributed to Pan-STARRS and ePESSTO+ data analysis. E.O.O. assisted with LAST follow-up observations and performed LAST data reduction and photometry. C.M.B.O. contributed to the source analysis. G.P. performed SMA follow-up observations and data analysis. A.R. performed CHIMERA observations and data reduction. R.R. created the SEDM2 robotic observing software. Y.S. performed KP84 observations. J.S. assisted with NOT follow-up observations. D.S. performed the GRB search. Y.Y. helped write the transient scanning code and assisted with X-ray observations.

 \item[Competing Interests] The authors declare no competing interests.

 \item[Correspondence] Correspondence and requests for materials
should be addressed to Anna Y. Q. Ho (email: annayqho@cornell.edu).

 \item[Data Availability] The reduced optical photometric data of \at\ are provided in Supplementary Table 1 and Supplementary Table 2. Spectroscopy of AT2022tsd will be made available via the WISeREP public database. Facilities that make all their data available in public archives, either promptly or after a proprietary period, include the VLA, the Liverpool Telescope, the W.~M. Keck Observatory, the Palomar 48-inch/ZTF,  the {\it Neil Gehrels Swift Observatory}, \emph{Chandra}, and ALMA. Additionally, all of the data required to reproduce the figures is available in a public Github repository (see ``Code Availability'').
 
 \item[Code Availability] The code and data used to perform the calculations and produce the figures for this paper are available in a public Github repository.\footnote{\url{www.github.com/annayqho/AT2022tsd}}

\end{addendum}

\begin{extended_data}

\renewcommand{\thefigure}{\arabic{figure}~Extended~Data}
\renewcommand{\thefigure}{Extended Data Figure \arabic{figure}}
\renewcommand{\figurename}{}
\setcounter{figure}{0}

\renewcommand{\thetable}{\arabic{table}~Extended~Data}
\renewcommand{\thetable}{Extended Data Table \arabic{table}}
\renewcommand{\tablename}{}
\setcounter{table}{0}

\begin{figure}[ht]
 \centering
\includegraphics[]{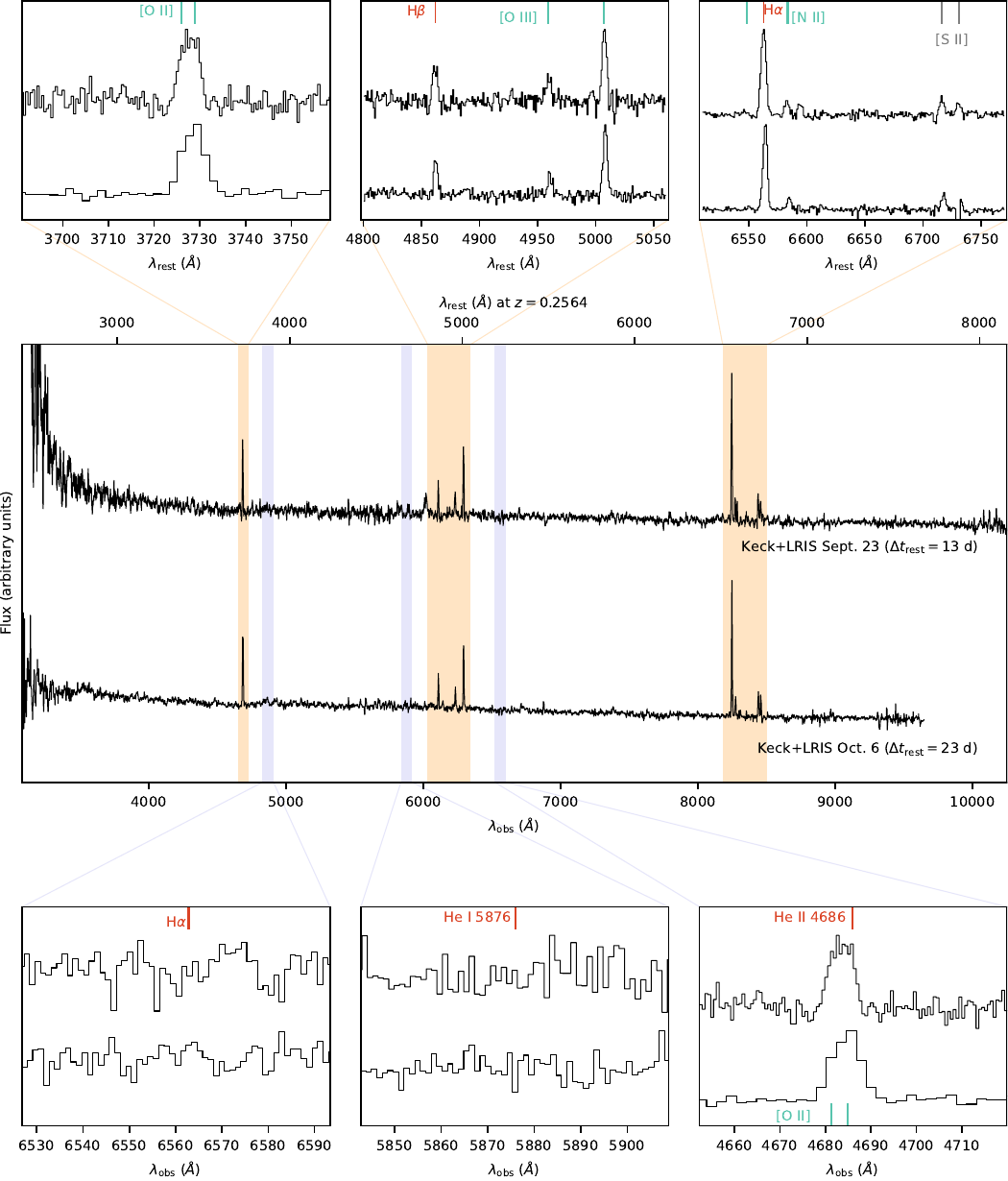}
  \caption{\textbf{Optical spectra of \at\ obtained with Keck/LRIS.} Spectra are binned using 3\AA\ bins. Regions with identified narrow host-galaxy emission lines, used to measure the best-fit redshift of $z=0.2564\pm0.0003$, are marked. Regions used to search for $z=0$ emission lines, as would be expected from a foreground Galactic transient, are also marked.}
 \label{fig:spec}
\end{figure}

\begin{figure}[!ht]
 \centering
 \begin{subfigure}[t]{\textwidth}
  	\centering
	\includegraphics[]{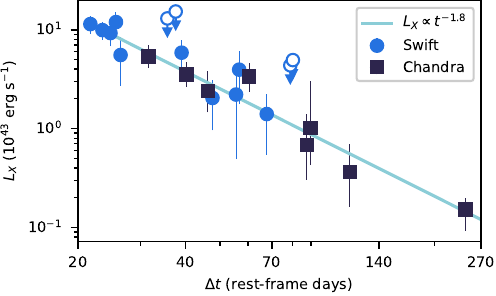}
	 \caption*{\textbf{a}}
 \end{subfigure}
  \begin{subfigure}[t]{\textwidth}
  	\centering
	 \includegraphics[]{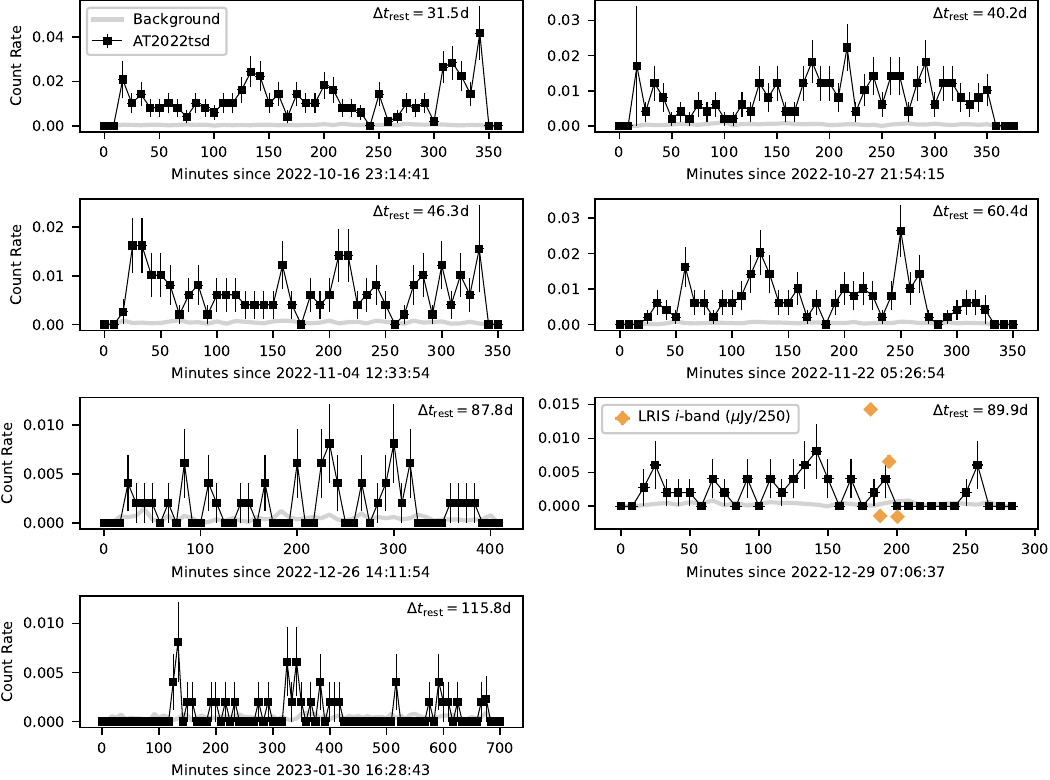}
	 \caption*{\textbf{b}}
 \end{subfigure}
\caption{\textbf{X-ray (0.3--10\,keV) light curve of \at.} \textbf{a} Full light curve with best-fit power law of $\alpha=-1.81\pm0.13$, where $f_\nu \propto t^{\alpha}$. Upper limits (3$\sigma$) are shown with open circles. \textbf{b} Individual {\it Chandra} observations binned in time with 500\,s bins. Diamonds show an optical ($i$-band) flare detected with LRIS during one of the {\it Chandra} observations. Error bars are 1$\sigma$ confidence intervals.}
 \label{fig:xray-lc}
\end{figure}

\begin{figure}[!ht]
 \centering
  \begin{subfigure}[t]{\textwidth}
  \centering
 \includegraphics[]{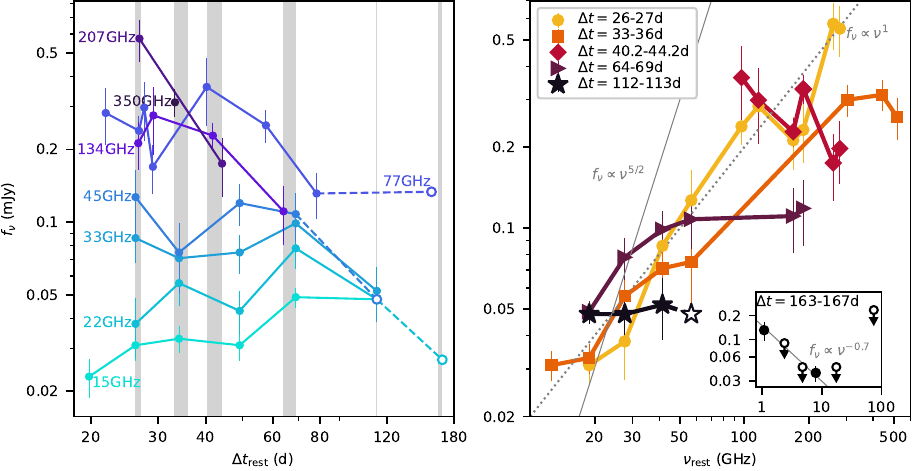}
 \caption*{\textbf{a}}
 \end{subfigure}
  \begin{subfigure}[t]{\textwidth}
  \centering
 \includegraphics[]{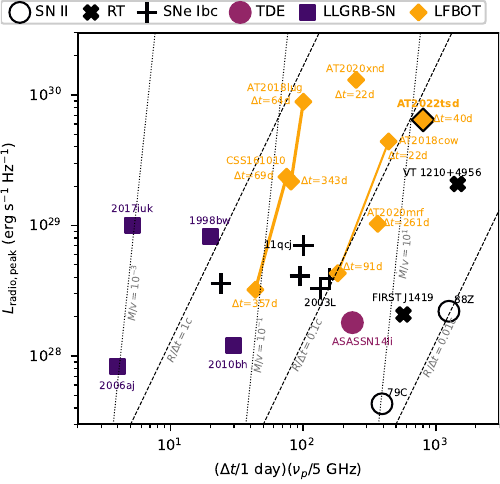}
 \caption*{\textbf{b}}
 \end{subfigure}
\caption{\textbf{Radio data of \at.} \textbf{a} Selected radio light curves and SEDs from the VLA (15--45\,GHz), NOEMA (77--207\,GHz), and ALMA ($350$\,GHz). Open circles mark 5$\sigma$ upper limits, and dashed lines connect upper limits to detections. Vertical shaded regions mark epochs of rest-frame radio SEDs. Inset shows SED from late-time observations with the GMRT and VLA. Solid line marks the $f_\nu \propto \nu^{5/2}$ power law expected from synchrotron self-absorption, and dotted line marks the shallower $f_\nu \propto \nu^{1}$. \textbf{b} Peak frequency ($\nu_p$) at a fixed time post-explosion ($\Delta t$) vs. peak luminosity of extragalactic radio transients. Error bars are 1$\sigma$ confidence intervals. See Methods section~\ref{sec:data-transient-parameter-space} for additional details and data sources.}
 \label{fig:radio}
\end{figure}

\clearpage

\begin{figure}[ht]
 \centering
\includegraphics[]{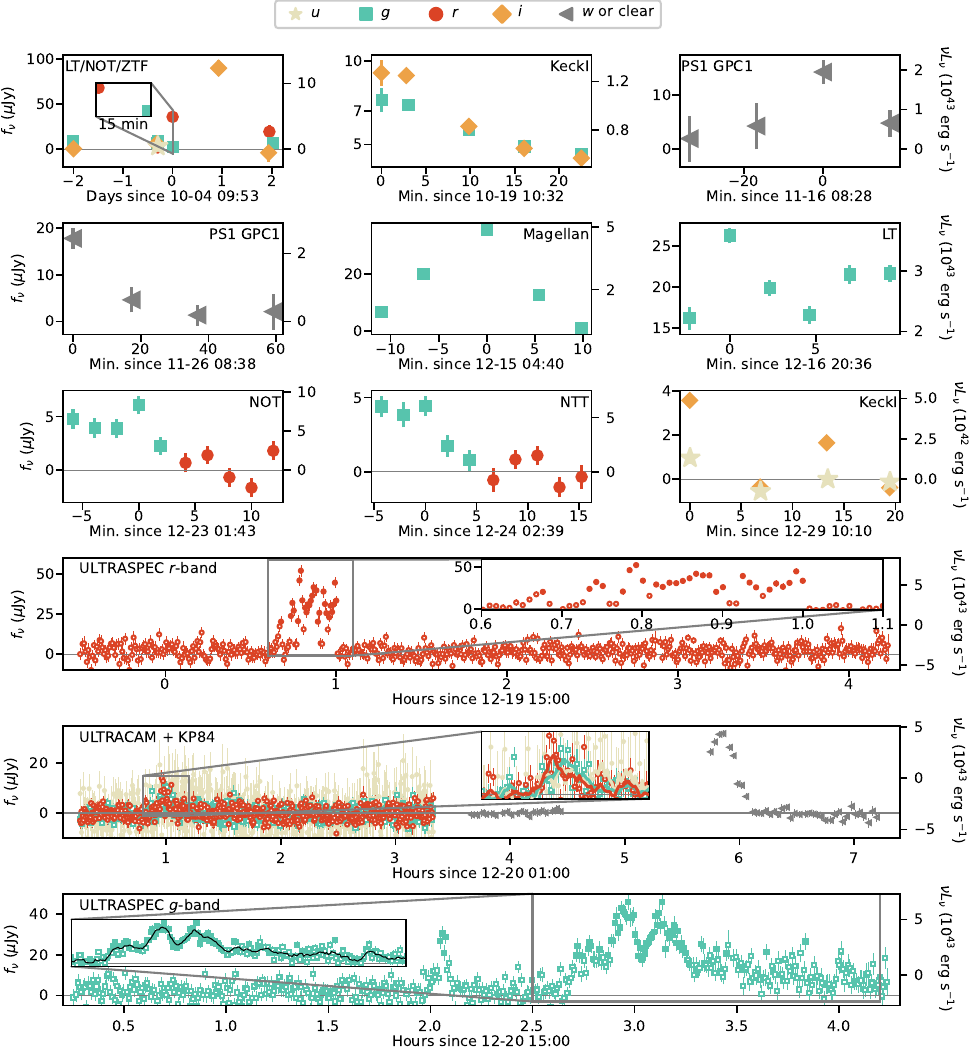}
  \caption{\textbf{Collage of \at\ flares, corrected for Milky Way extinction.} For ULTRASPEC, ULTRACAM, and KP84, open points are $<5\sigma$ and filled points are $\geq5\sigma$. The insets of the ULTRASPEC and ULTRACAM light curves show 3\,min and 1\,min running averages, respectively. Error bars are 1$\sigma$ confidence intervals.
The cadence of the ULTRASPEC, ULTRACAM, and KP84 light curves are 30\,s, 20\,s, and 120\,s respectively, and the filled points are 
}
 \label{fig:flare-collage}
\end{figure}

\begin{figure*}[ht]
    \centering
    \includegraphics[]{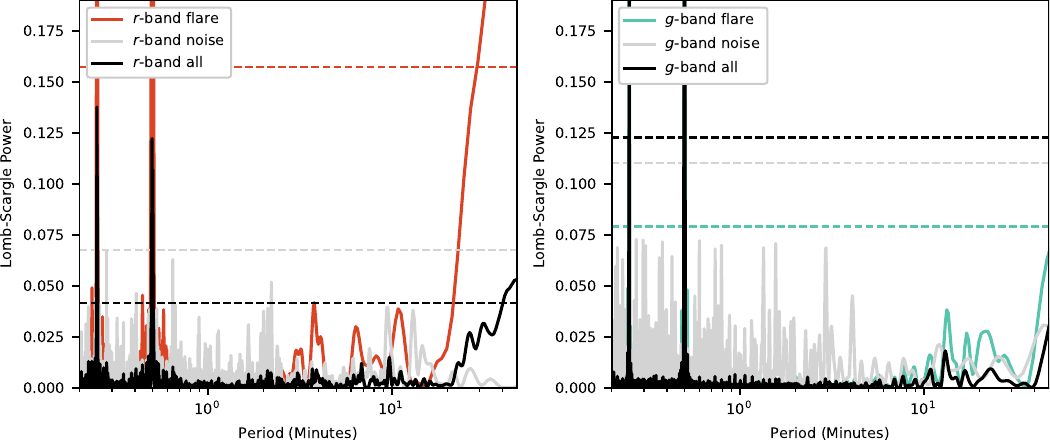}
    \caption{\textbf{Lomb-Scargle periodogram of the ULTRASPEC flares.} Each panel shows the periodogram for the flare itself, for a region of the light curve with no significant detections (``noise"), and for the full light curve (``all"). Horizontal dashed lines mark the power expected for a false-alarm peak (with false-alarm probability 2.5\%) under the assumption that there is no periodicity present in the data, using a bootstrap simulation. The only peaks higher than this threshold are from the cadence of the observation (30\,s, and an alias at half that value), from the overall flare width, and from the duration of the observation.}
    \label{fig:ultraspec-periodogram}
\end{figure*}

\begin{figure*}[!ht]
    \centering
    \includegraphics[]{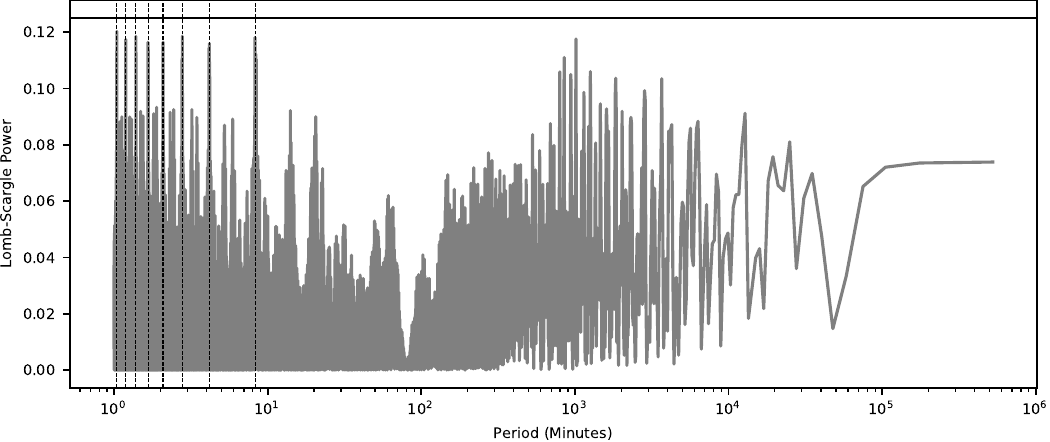}
    \caption{\textbf{Lomb-Scargle periodogram of X-ray observations.}
    Periodogram was constructed using the first four epochs of {\it Chandra} X-ray data. The horizontal line shows the power expected for a false-alarm peak (with false-alarm probability 2.5\%) under the assumption that there is no periodicity present in the data, using a bootstrap simulation. The observed peaks arise from the 500\,s sampling and aliases (marked with vertical dotted lines).}
    \label{fig:xray-periodogram}
\end{figure*}

\begin{figure*}[!ht]
    \centering
    \includegraphics[]{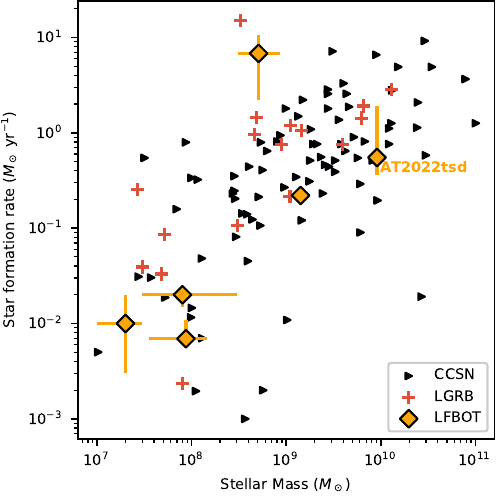}
    \caption{\textbf{The stellar mass and star-formation rate (SFR) of \at's host galaxy.} For reference we show the stellar mass and star-formation rate of other transient host galaxies\cite{Taggart2021}, including core-collapse supernovae\cite{Taggart2021}, long-duration $\gamma$-ray bursts\cite{Taggart2021}, and luminous fast blue optical transients\cite{Perley2019,Ho2020_Koala,Coppejans2020,Perley2021,Yao2022}. Error bars are 1$\sigma$ confidence intervals.}
    \label{fig:host-galaxy}
\end{figure*}

\clearpage

\begin{table}
\begin{center}
\caption{\textbf{Summary of targeted flare searches.} The number of exposures $N_\mathrm{exp}$, the total observing time $T_\mathrm{exp}$, the typical depth per exposure, and the number of flares detected are listed. For reference, we include the Magellan/IMACS observation in which flaring was first noticed.}
\label{tab:flare-searches}
 \begin{tabular}{lcccccc} 
 \hline\hline
 Telescope & Filters & $N_\mathrm{exp}$ & $T_\mathrm{exp}$ (min) & Depth (AB mag) & \# Flares \\
 \hline
 Magellan/IMACS & $g$ & 4 & 12 & 24.2 & 1 \\
 LT/IO:O & $gr$ & 134 & 265 & 22.6 & 1 \\
 NOT/ALFOSC & $gr$ & 15 & 20 & 23.5 & 1 \\
 NTT/ULTRACAM & $giru$ & 1981 & 660 & 22.3 & 1 \\
 TNT/ULTRASPEC & $gr$ & 1045 & 519 & 22.0 & 3 \\
 KP84/SEDM2 & clear & 60 & 120 & 22.7 & 1 \\
 NTT/EFOSC & $gr$ & 30 & 47 & 23.7 & 1 \\
 GIT & clear & 59 & 295 & 21.1 & 0 \\
 HCT & $R$ & 55 & 275 & 22.4 & 0 \\
 SLT & $r$ & 28 & 140 & 99.0 & 0 \\
 LOT & $g$ & 27 & 135 & 99.0 & 0 \\
 KeckI/LRIS & $giu$ & 16 & 71 & 24.8 & 2 \\
 %PS1 GPC1 & $i$ & 8 & 6 & 20.6 & 2 \\
 P200/CHIMERA & $gr$ & 420 & 350 & 21.3 & 0 \\
 LAST & $G_p$ & 646 & 9312 & 20.0 & 0 \\
 \hline
 \end{tabular}
 \end{center}
\end{table}

\clearpage

\begin{table}
\begin{center}
\caption{\textbf{Summary of \at\ flare properties.} We include time of brightest detection ($t_\mathrm{peak,obs}$), time interval in which 90\% of the flux was measured ($T_{90}$), peak luminosity ($\nu L_\nu$ in the specified band), and total energy radiated $E_\mathrm{rad}$. Flares are defined as $\geq 5\sigma$ detections, verified visually, with an MJD after 59856.4 ($\Delta t_\mathrm{obs}=27\,$d). In cases with flares observed in multiple filters, quantities are calculated using the first filter listed. Note that, with the exception of the ULTRASPEC and ULTRACAM sequences, observations did not capture the start and end of the flare.}
\label{tab:flare-properties}
 \begin{tabular}{cccccc} 
 \hline\hline
 $t_\mathrm{peak,obs}$ (MJD) & Telescope & Band & $T_{90,\mathrm{obs}}$ (min) & $L_\mathrm{peak,obs}$ (erg\,s$^{-1}$) & $E_\mathrm{rad}$ (erg) \\
 \hline
 59856.4122 & P48/ZTF & $r$ & -- & $>4\times10^{43}$ & -- \\
 59857.3403 & P48/ZTF & $i$ & -- & $>8\times10^{43}$ & -- \\
 59871.4392 & Keck1/LRIS & $gi$ & $>20$ & $>1\times10^{43}$ &  $>2\times10^{46}$ \\
 59899.3533 & PS1/GPC1 & $w$ & 40 & $2\times10^{43}$ & $4\times10^{46}$ \\ 
 59909.3598 & PS1/GPC1 & $w$ & $>50$ & $>2\times10^{43}$ & $>6\times10^{46}$ \\
 59928.1951 & Magellan/IMACS & $g$ & 16 & $6\times10^{43}$ & $6\times10^{46}$ \\
 59929.8585 & LT/IO:O & $g$ & 10 & $4\times10^{43}$ & $2\times10^{46}$ \\ 
 59932.6580 & TNT/ULTRASPEC & $r$ & 19 & $5\times10^{43}$ & $6\times10^{46}$ \\
 59933.0822 & NTT/ULTRACAM & $rgu$ & 12 & $8\times10^{42}$ & $3\times10^{45}$  \\
 59933.2858 & KP84/SEDM2 & clear & $>15$ & $2\times10^{43}$ & $>2\times10^{46}$  \\
 59933.7107 & TNT/ULTRASPEC & $g$ & 7 & $2\times10^{43}$ & $8\times10^{45}$ \\
 59933.7556 & TNT/ULTRASPEC & $g$& 78 & $3\times10^{43}$ & $1\times10^{47}$  \\
 59936.0720 & NOT/ALFOSC & $g$ & $>15$ & $>8\times10^{42}$ & $3\times10^{45}$ \\
 59937.1105 & NTT/EFOSC & $g$ & $>8$ & $>6\times10^{42}$ & $2\times10^{45}$ \\
 59942.4238 & Keck1/LRIS & $iu$ & -- & $>3\times10^{42}$ & -- \\
 \hline
 \end{tabular}
 \end{center}
\end{table}

\clearpage

\begin{table}
\begin{center}
\caption{\textbf{\at\ flare duty cycle for different apparent-magnitude thresholds.} We consider the date range MJD 59,856.41--59,942.43 (from the first flare detection to the last flare detection). $N_\mathrm{exp}$ is the number of exposures brighter than the given magnitude threshold, $T_\mathrm{exp}$ is the total exposure time, $T_\mathrm{on}$ is the total time with a flare detected, and the bounds are 97.5\% confidence intervals (see Methods section~\ref{sec:flare-characteristics}) on the duty cycle $T_\mathrm{on}/T_\mathrm{exp}$. }
\label{tab:flare-stats}
 \begin{tabular}{ccccc} 
 \hline\hline
 Threshold (AB Mag) & $N_\mathrm{exp}$ & $T_\mathrm{exp}$ (Minutes) & $T_\mathrm{on}$/$T_\mathrm{exp}$ & Bounds \\
 \hline
 21.0 & 1271 & 1142 & 0.02 & [0.001, 0.1] \\
 22.5 & 68 & 155 & 0.1 & [0.01, 0.6]  \\
 24.0 & 13 & 65 & 0.5 & [0.03, 1] \\
 \hline
 \end{tabular}
 \end{center}
\end{table}

\end{extended_data}

\clearpage

\begin{supplement}

\renewcommand{\thefigure}{Supplementary Information Figure~\arabic{figure}}
\renewcommand{\figurename}{}
\setcounter{figure}{0}

\renewcommand{\thetable}{Supplementary Information Table~\arabic{table}}
 \renewcommand{\tablename}{}
\setcounter{table}{0}

\begin{table}
    \centering
    \begin{tabular}{llllll}
    \hline\hline
     Object & Band & $L_\mathrm{flare}$ (erg\,s$^{-1}$) & Amp. & Duration & Persistence \\
     \hline
     \multicolumn{6}{c}{\emph{Unknown}}\\
    AT2022tsd (this paper) & 500\,nm & $10^{43}$--$10^{44}$ & $\gtrsim100\times$ & 10--80\,min & $\gtrsim100$\,d \\
    GRB\,070610 (BH? NS?) & 800\,nm & $10^{35}$? & $\gtrsim100\times$ & 10\,s--mins & 5\,d \\
     NGC 1313 X-2 (ULX) & 0.3--10\,keV & $10^{40}$ & $\sim10\times$ & 10\,min & -- \\
     \multicolumn{6}{c}{\emph{Neutron Stars}}\\
    SGR in M81/M82 (GF Spike) & 20\,keV--10\,MeV & $1.8\times10^{47}$ & $\sim 10^{11}\times$ & 0.5\,s & -- \\
    SGR 1806-20 (GF Tail)  & 20\,keV--10\,MeV & $1.3\times10^{42}$ & $\sim 10^{7}\times$ & 8\,min & -- \\
    Crab (nanoshot) & 8\,GHz & $10^{34}$ & $>1000\times$ & 2\,ns & -- \\
     \multicolumn{6}{c}{\emph{Stellar-mass black holes}}\\
    GRS 1915+105 (XRB) & 2.2\,$\mu$m & $\gtrsim10^{36}$ & $\lesssim 10\times$ & 10\,min & -- \\
    GRB\,080319B  (GRB) & 500\,nm & $10^{50}$ & $>10\times$ & 40\,s & 60\,s \\
     \multicolumn{6}{c}{\emph{Supermassive black holes}}\\
    AT2019ehz (TDE) & 0.3--10\,keV & $10^{44}$ & $>10\times$ & 10\,d & 70\,d \\  
    Sagittarius A* & 2.1\,$\mu$m & $10^{34}$ & $\lesssim 10\times$ & 30\,min & -- \\
    M87 & 350\,GeV & $10^{42}$ & $\gtrsim10\times$ & Few days & -- \\
    S5 1803+784 (blazar) & 600\,nm & $10^{46}$ & $10\times$ & $\gtrsim1\,$month & -- \\
    GSN\,069 (QPE) & 0.4--1\,keV & $10^{43}$ & $\gtrsim10\times$ & 1\,hr & -- \\
    ASASSN-14ko (TDE?) & 200--500\,nm & $10^{43}$--$10^{44}$ & $>10\times$ & 10\,d & -- \\
    \hline\hline
    \end{tabular}
    \caption{\textbf{Summary of large-amplitude ($\gtrsim10\times$) flares from representative literature objects.} See Methods section~\ref{sec:data-table} for additional details and data sources.}
    \label{tab:flaring-classes}
\end{table}

\clearpage

\begin{table}
\begin{center}
\caption{\textbf{Host-galaxy photometry for \at, not corrected for Milky Way extinction.} Error bars are 1$\sigma$ confidence intervals.}
\label{tab:host-photometry}
\begin{tabular}{ccc} 
\hline\hline
Survey & Filter & Brightness (AB mag) \\
\hline
PanSTARRS & $g$               &$ 21.32 \pm 0.10 $\\
PanSTARRS & $r$               &$ 20.59 \pm 0.07 $\\
PanSTARRS & $i$               &$ 20.67 \pm 0.05 $\\
PanSTARRS & $z$               &$ 20.87 \pm 0.36 $\\
PanSTARRS & $y$               &$ 20.14 \pm 0.10 $\\
\hline
\end{tabular}
\end{center}
\end{table}

\begin{figure}[!ht]
 \centering
\includegraphics[width=0.6\textwidth]{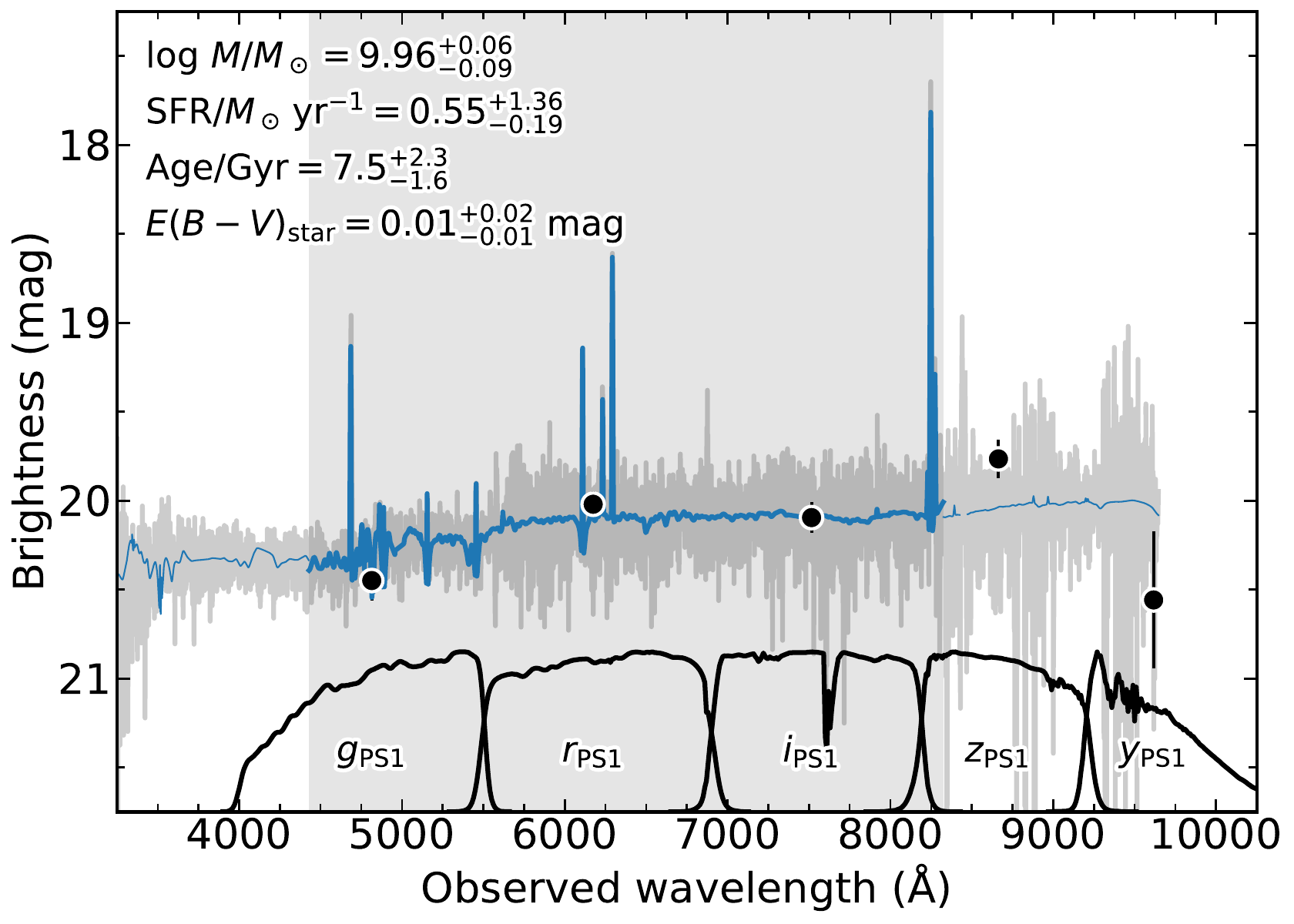}
  \caption{\textbf{Best-fit host-galaxy properties of \at.} The observed host-galaxy photometry is shown as black data points, the observed host-galaxy spectrum is shown in grey, and the best-fit model is shown in blue. The shaded region indicates the region of the spectrum used in the \texttt{prospector} fit.
}
 \label{fig:host-fit}
\end{figure}

\begin{figure}[!ht]
 \centering
\includegraphics[]{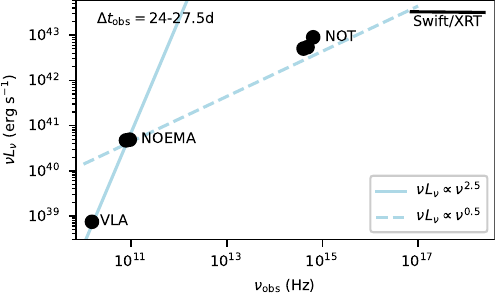}
  \caption{\textbf{SED of \at\ at $\Delta t_\mathrm{obs}\approx25\,$d post-discovery.} X-ray data are shown with a photon index of $\Gamma=2.01$ across the {\it Swift}/XRT 0.3--10\,keV bandpass. Lines mark power laws connecting the radio to submillimeter data (solid), and the millimeter to X-ray data (dashed).}
 \label{fig:full-sed}
\end{figure}

\begin{table*} 
\centering 
\begin{tabular}{llllll}
\hline\hline
$t$ & $\Delta t$ & $t_\mathrm{exp}$ & Count Rate & $F_X$ & $L_X$ \\ 
(UT) & (days) & (ks) & ($10^{-3}$\,s$^{-1}$) & ($10^{-14}$ erg\,s$^{-1}$\,cm$^{-2}$) & $(10^{43}$\,erg\,s$^{-1}$) \\ 
\hline
2022-10-04 09:17 & $22.65\pm0.24$ & 3.64 & $10.43\pm2.06$ & $53.17\pm10.50$ & $11.43\pm2.26$  \\ 
2022-10-06 14:55 & $24.41\pm0.22$ & 3.78 & $9.06\pm1.85$ & $46.19\pm9.44$ & $9.93\pm2.03$  \\ 
2022-10-08 02:17 & $25.65\pm0.29$ & 2.47 & $8.46\pm2.24$ & $43.14\pm11.43$ & $9.28\pm2.46$  \\ 
2022-10-09 05:06 & $26.54\pm0.29$ & 2.29 & $10.92\pm2.91$ & $55.67\pm14.84$ & $11.97\pm3.19$  \\ 
2022-10-10 09:47 & $27.31\pm0.11$ & 2.37 & $5.07\pm2.60$ & $25.85\pm13.26$ & $5.56\pm2.85$  \\ 
2022-10-21 16:35 & $36.60\pm0.42$ & 1.44 & $<11.89$ & $<60.63$ & $<13.04$  \\ 
2022-10-24 09:25 & $38.57\pm0.24$ & 1.04 & $<13.97$ & $<71.27$ & $<15.32$  \\ 
2022-10-26 01:27 & $40.03\pm0.37$ & 2.77 & $5.35\pm1.88$ & $27.30\pm9.56$ & $5.87\pm2.06$  \\ 
2022-11-06 01:21 & $48.65\pm0.24$ & 4.39 & $1.86\pm0.99$ & $9.50\pm5.04$ & $2.04\pm1.08$  \\ 
2022-11-16 01:40 & $56.48\pm0.11$ & 1.87 & $2.02\pm1.57$ & $10.29\pm8.01$ & $2.21\pm1.72$  \\ 
2022-11-17 07:44 & $57.61\pm0.24$ & 1.96 & $3.60\pm1.99$ & $18.38\pm10.16$ & $3.95\pm2.18$  \\ 
2022-12-01 02:23 & $68.65\pm0.32$ & 5.75 & $1.28\pm0.78$ & $6.54\pm3.99$ & $1.41\pm0.86$  \\ 
2022-12-15 00:09 & $79.78\pm0.38$ & 2.97 & $<3.99$ & $<20.33$ & $<4.37$  \\ 
2022-12-16 09:52 & $81.10\pm0.58$ & 2.67 & $<4.50$ & $<22.97$ & $<4.94$  \\ 
\hline 
\end{tabular} 
\caption{\textbf{{\it Swift} XRT (0.3--10\,keV) observations of \at.} The table includes epochs $\Delta t$ since discovery in the rest frame, exposure time $t_\mathrm{exp}$, flux $F_X$, and luminosity $L_X$. Error bars are 1$\sigma$ and upper limits are given as 3$\sigma$.} 
\label{tab:xray-observations} 
\end{table*} 

\begin{table*} 
\centering 
\begin{tabular}{lllll}
\hline\hline
$t_\mathrm{start}$ & $\Delta t$ & $t_\mathrm{exp}$ & $F_X$ & $L_X$\\ 
(UT) & (days) & (ks) & ($10^{-14}$ erg\,s$^{-1}$\,cm$^{-2}$) & $(10^{43}$\,erg\,s$^{-1}$)\\ 
\hline
2022-10-16 23:14 & 32.42 & 20 & $14.60^{+3.33}_{-3.22}$ & $3.14^{+0.72}_{-0.69}$ \\ 
2022-10-27 21:54 & 41.13 & 20 & $10.46^{+2.78}_{-2.22}$ & $2.25^{+0.60}_{-0.48}$ \\ 
2022-11-04 12:33 & 47.19 & 20 & $7.59^{+2.64}_{-2.40}$ & $1.63^{+0.57}_{-0.52}$ \\ 
2022-11-22 05:26 & 61.27 & 20 & $9.17^{+3.14}_{-2.53}$ & $1.97^{+0.68}_{-0.54}$ \\ 
2022-12-26 14:11 & 88.62 & 24 & $1.68^{+2.03}_{-0.92}$ & $0.36^{+0.44}_{-0.20}$ \\ 
2022-12-29 07:06 & 90.77 & 16 & $2.48^{+4.98}_{-1.57}$ & $0.53^{+1.07}_{-0.34}$ \\ 
2023-01-30 16:28 & 116.55 & 40 & $0.96^{+1.04}_{-0.51}$ & $0.21^{+0.22}_{-0.11}$ \\ 
2023-07-11 03:37 & 244.09 & 16 & $0.66^{+0.54}_{-0.43}$ & $0.14^{+0.12}_{-0.09}$ \\ 
2023-07-11 to 2023-07-16 & 244--248 & 40 & $0.38^{+0.15}_{-0.12}$ & $0.08^{+0.03}_{-0.03}$ \\ 
\hline 
\end{tabular} 
\caption{\textbf{{\it Chandra X-ray Observatory} 0.5--6\,keV observations of AT2022tsd.} The table includes epochs $\Delta t$ since discovery in the rest frame, exposure time $t_\mathrm{exp}$, flux $F_X$, and luminosity $L_X$. Error bars are 1$\sigma$ confidence intervals. The final row shows the stacked measurement from three observations conducted on three different days.} 
\label{tab:chandra} 
\end{table*}

\clearpage

\begin{center} 
\begin{longtable}{llllll} 
\caption{\textbf{Radio observations of AT2022tsd.} Table includes epochs since discovery $\Delta t$ in the rest frame, observed frequency $\nu_\mathrm{obs}$, flux density $f_\nu$ of the source (if detected), and root-mean-square (RMS) of a region close to the source in the image.} 
\label{tab:radio-observations}\\ 
\hline\hline
Start Date & $\Delta t$ & $\nu_\mathrm{obs}$ & $f_\nu$ & RMS & Telescope\\ 
(UT) & (days) & (GHz) & (mJy) & (mJy) & \\ 
\hline
2022-10-02 06:50:00 & 19.75 & 15.00 & $0.023$ & $0.004$ & VLA \\ 
2022-10-04 07:20:00 & 21.36 & 230.00 & -- & $0.270$ & SMA \\ 
2022-10-04 22:07:00 & 21.85 & 77.26 & $0.283$ & $0.075$ & NOEMA \\ 
2022-10-04 22:07:00 & 21.85 & 92.74 & $0.245$ & $0.065$ & NOEMA \\ 
2022-10-10 08:02:00 & 26.16 & 45.00 & $0.127$ & $0.033$ & VLA \\ 
2022-10-10 08:02:00 & 26.16 & 22.00 & $0.038$ & $0.009$ & VLA \\ 
2022-10-10 08:02:00 & 26.16 & 15.00 & $0.031$ & $0.004$ & VLA \\ 
2022-10-10 08:02:00 & 26.16 & 33.00 & $0.086$ & $0.013$ & VLA \\ 
2022-10-10 21:16:00 & 26.59 & 134.76 & $0.212$ & $0.047$ & NOEMA \\ 
2022-10-10 21:16:00 & 26.59 & 150.24 & $0.232$ & $0.057$ & NOEMA \\ 
2022-10-11 00:45:00 & 26.71 & 77.26 & $0.239$ & $0.035$ & NOEMA \\ 
2022-10-11 00:45:00 & 26.71 & 92.74 & $0.284$ & $0.032$ & NOEMA \\ 
2022-10-11 02:53:00 & 26.78 & 207.26 & $0.574$ & $0.114$ & NOEMA \\ 
2022-10-11 02:53:00 & 26.78 & 222.74 & $0.551$ & $0.117$ & NOEMA \\ 
2022-10-12 02:50:00 & 27.57 & 77.26 & $0.298$ & $0.082$ & NOEMA \\ 
2022-10-12 02:50:00 & 27.57 & 92.74 & $0.316$ & $0.078$ & NOEMA \\ 
2022-10-13 23:24:00 & 29.05 & 77.26 & $0.170$ & $0.039$ & NOEMA \\ 
2022-10-13 23:24:00 & 29.05 & 92.74 & $0.179$ & $0.037$ & NOEMA \\ 
2022-10-14 02:04:00 & 29.14 & 134.76 & $0.277$ & $0.087$ & NOEMA \\ 
2022-10-14 02:04:00 & 29.14 & 150.24 & $0.411$ & $0.117$ & NOEMA \\ 
2022-10-19 04:29:00 & 33.20 & 350.50 & $0.313$ & $0.027$ & ALMA \\ 
2022-10-20 05:44:00 & 34.04 & 15.00 & $0.033$ & $0.004$ & VLA \\ 
2022-10-20 05:44:00 & 34.04 & 33.00 & $0.071$ & $0.010$ & VLA \\ 
2022-10-20 05:44:00 & 34.04 & 22.00 & $0.056$ & $0.007$ & VLA \\ 
2022-10-20 05:44:00 & 34.04 & 10.00 & $0.031$ & $0.004$ & VLA \\ 
2022-10-20 05:44:00 & 34.04 & 45.00 & $0.075$ & $0.021$ & VLA \\ 
2022-10-21 04:54:40 & 34.81 & 412.00 & $0.259$ & $0.038$ & ALMA \\ 
2022-10-22 03:52:39 & 35.57 & 242.00 & $0.300$ & $0.028$ & ALMA \\ 
2022-10-28 00:54:00 & 40.25 & 77.26 & $0.363$ & $0.113$ & NOEMA \\ 
2022-10-28 00:54:00 & 40.25 & 92.74 & $0.299$ & $0.093$ & NOEMA \\ 
2022-10-28 00:54:00 & 40.25 & 150.24 & $0.328$ & $0.037$ & NOEMA \\ 
2022-10-29 23:00:00 & 41.77 & 150.24 & $0.330$ & $0.040$ & NOEMA \\ 
2022-10-29 23:00:00 & 41.77 & 134.76 & $0.228$ & $0.028$ & NOEMA \\ 
2022-11-01 23:03:00 & 44.16 & 222.74 & $0.198$ & $0.052$ & NOEMA \\ 
2022-11-01 23:03:00 & 44.16 & 207.26 & $0.175$ & $0.048$ & NOEMA \\ 
2022-11-08 04:52:00 & 49.13 & 15.00 & $0.031$ & $0.004$ & VLA \\ 
2022-11-08 04:52:00 & 49.13 & 22.00 & $0.043$ & $0.006$ & VLA \\ 
2022-11-08 04:52:00 & 49.13 & 33.00 & $0.075$ & $0.008$ & VLA \\ 
2022-11-08 04:52:00 & 49.13 & 45.00 & $0.120$ & $0.015$ & VLA \\ 
2022-11-18 20:08:00 & 57.60 & 77.26 & $0.252$ & $0.039$ & NOEMA \\ 
2022-11-18 20:08:00 & 57.60 & 92.74 & $0.304$ & $0.030$ & NOEMA \\ 
2022-11-26 22:16:00 & 64.04 & 134.76 & $0.111$ & $0.030$ & NOEMA \\ 
2022-11-26 22:16:00 & 64.04 & 150.24 & $0.119$ & $0.032$ & NOEMA \\ 
2022-12-03 03:26:00 & 68.98 & 22.00 & $0.078$ & $0.007$ & VLA \\ 
2022-12-03 03:26:00 & 68.98 & 33.00 & $0.099$ & $0.009$ & VLA \\ 
2022-12-03 03:26:00 & 68.98 & 45.00 & $0.108$ & $0.018$ & VLA \\ 
2022-12-03 03:26:00 & 68.98 & 15.00 & $0.049$ & $0.004$ & VLA \\ 
2022-12-14 18:56:00 & 78.25 & 77.25 & $0.131$ & $0.028$ & NOEMA \\ 
2022-12-14 18:56:00 & 78.25 & 92.74 & $0.153$ & $0.024$ & NOEMA \\ 
2023-01-27 01:26:00 & 112.69 & 45.00 & -- & $0.016$ & VLA \\ 
2023-01-27 01:26:00 & 112.69 & 33.00 & $0.052$ & $0.011$ & VLA \\ 
2023-01-27 01:26:00 & 112.69 & 15.00 & $0.048$ & $0.003$ & VLA \\ 
2023-01-27 01:26:00 & 112.69 & 22.00 & $0.048$ & $0.006$ & VLA \\ 
2023-03-04 12:14 & 141.70 & 1.27 & $0.140$ & $0.033$ & uGMRT \\ 
2023-03-05 12:14 & 142.50 & 0.65 & -- & $0.195$ & uGMRT \\ 
2023-03-06 10:19 & 143.23 & 0.44 & -- & $0.810$ & uGMRT \\ 
2023-03-23 13:19:00 & 156.86 & 77.25 & -- & $0.045$ & NOEMA \\ 
2023-03-23 13:19:00 & 156.86 & 92.74 & -- & $0.047$ & NOEMA \\ 
2023-03-31 08:10 & 163.06 & 1.37 & $0.131$ & $0.035$ & uGMRT \\ 
2023-04-01 10:05 & 163.92 & 0.65 & -- & $0.165$ & uGMRT \\ 
2023-04-02 10:05 & 164.71 & 0.43 & -- & $0.465$ & uGMRT \\ 
2023-04-05 23:00:00 & 167.53 & 6.00 & -- & $0.009$ & VLA \\ 
2023-04-05 23:00:00 & 167.53 & 10.00 & $0.038$ & $0.009$ & VLA \\ 
2023-04-05 23:00:00 & 167.53 & 22.00 & -- & $0.009$ & VLA \\ 
2023-04-05 23:00:00 & 167.53 & 3.00 & -- & $0.018$ & VLA \\ 
\hline 
\end{longtable} 
\end{center}

\end{supplement}

\end{document}